\documentclass{article}
\usepackage[utf8]{inputenc}

\usepackage{geometry}
\usepackage{amsmath}
\allowdisplaybreaks
\usepackage{mathtools}
\usepackage{blkarray}
\usepackage{amssymb}
\usepackage{amsthm}
\usepackage[shortlabels]{enumitem}
\usepackage{setspace}
\usepackage[colorlinks,linkcolor=blue,citecolor=blue,urlcolor=blue,bookmarks=false,hypertexnames=true]{hyperref} 
\usepackage{xcolor}
\usepackage{bbm}

\usepackage[round]{natbib}
\newtheorem{theorem}{Theorem}
\numberwithin{theorem}{section}
\newtheorem{definition}[theorem]{Definition}
\newtheorem{axiom}[theorem]{Axiom}
\newtheorem{proposition}[theorem]{Proposition}
\newtheorem{lemma}[theorem]{Lemma}
\newtheorem{corollary}[theorem]{Corollary}
\newtheorem{example}[theorem]{Example}
\newtheorem{claim}[theorem]{Claim}

\newcommand{\R}{\mathbb{R}}
\newcommand{\mbb}{\mathbb}

\makeatletter
\def\@fnsymbol#1{\ensuremath{\ifcase#1\or \dagger\or \ddagger\or
   \mathsection\or \mathparagraph\or \|\or **\or \dagger\dagger
   \or \ddagger\ddagger \else\@ctrerr\fi}}
    \makeatother

\title{Dynamic Random Choice}
\author{Ricky Li\thanks{rickyli@mit.edu. I thank Victor Aguiar and Christopher Turansick for helpful comments. I especially thank Tomasz Strzalecki for his insightful comments, as well as his invaluable guidance throughout my research career.}}
\date{This version: June 20, 2022}

\begin{document}

\maketitle

\sloppy

\begin{abstract}
    I study dynamic random utility with finite choice sets and exogenous total menu variation, which I refer to as stochastic utility (SU). First, I characterize SU when each choice set has three elements. Next, I prove several mathematical identities for joint, marginal, and conditional Block--Marschak sums, which I use to obtain two characterizations of SU when each choice set but the last has three elements. As a corollary under the same cardinality restrictions, I sharpen an axiom to obtain a characterization of SU with full support over preference tuples. I conclude by characterizing SU without cardinality restrictions. All of my results hold over an arbitrary finite discrete time horizon.
\end{abstract}

\section{Introduction and Related Literature}
A classic result in decision theory is \cite{sen1971choice}'s characterization of deterministic choice functions that can be represented by strict preference relations. However, economic choice data is often nondeterministic. In such cases, the analogous primitive and representation is a \textit{stochastic} choice function (SCF) and \textit{random} utility (RU) model. \cite{block1959random} define RU on arbitrary finite choice sets and show that their axiom requiring that the SCF's \textit{Block--Marschak sums} be nonnegative is necessary; \cite{falmagne1978representation} and \cite{barbera1986falmagne} complete the characterization of RU by showing that this axiom is sufficient.

In addition to nondeterministic choices, economic agents also often make choices across time, leading to a richer primitive than that of static random environments: a \textit{dynamic} stochastic choice function (DSCF). The corresponding economic model of interest is dynamic random utility (DRU), some variants of which have been examined in recent work. \cite{frick2019dynamic} examine dynamic random choice over the domain of \cite{kreps1978temporal}'s decision trees, where choices in each period are lotteries over consumption-continuation menu pairs. In this choice environment, choices are risky and future menu variation can be endogenously determined. Furthermore, since present choices constrain the set of possible future menus, the DSCF exhibits what \cite{frick2019dynamic} deem \textit{limited observability}. \cite{frick2019dynamic} characterize a model of dynamic random expected utility over arbitrary finite time periods, as well as sharper Bayesian variants with nonmyopic agents. \cite{kashaev2022nonparametric} axiomatize a model of DRU over a domain of consumption vectors in $\R_+^K$. In their setup, each period's menus are derived from exogenous budget planes, following the static framework of \cite{kitamura2018nonparametric}. Hence, \cite{kashaev2022nonparametric}'s axioms exploit \textit{partial} exogenous menu variation by assuming that only a finite subset of the set of menus that can be derived from arbitrary subsets of $\R_+^k$ are observable. Unlike the aforementioned choice environments, I study a model of DRU with riskless finite choice sets and full exogenous menu variation. To distinguish this setting from the others, I will henceforth refer to this paper's variant of DRU as \textit{stochastic utility} (SU), as originally named and defined in \cite{SC}.

The recent work most similar to mine is \cite{chambers2021correlated}, who study a model of correlated random utility (CRU) with riskless finite choice sets and full exogenous menu variation. Their primitive is a (two-agent) \textit{correlated choice rule}, whereas my primitive is a (multi-period) DSCF. In this paper, I show that under arbitrary extrapolation following zero-probability choices, each DSCF induces a stream of multi-agent correlated choice rules, each of which is marginally consistent in the last period. I also show that there is an equivalence between SU representations of a DSCF and CRU representations of the induced full-period correlated choice rule. One interpretation that bridges the gap between our models is to imagine the agent's current and future selves in my setup as different agents in \cite{chambers2021correlated}'s setup.

Using a graph-theoretic approach, \cite{chambers2021correlated} characterize CRU with two agents where at least one choice set is three or less elements. I provide two characterizations of SU for arbitrary finite time periods where all but the last choice set is three elements. One of those characterizations leverages axioms whose two-agent analogs play a similar role in the aforementioned result of \cite{chambers2021correlated}, but my approach uses a different proof strategy that does not involve graphs. I discuss the relationship between my and their axioms in greater detail later in this paper. \cite{chambers2021correlated} also find a counterexample to CRU under their axioms for larger choice environments, state an additional axiom which disciplines the correlated choice rule's capacity, and characterize CRU without cardinality restrictions using an analog of \cite{mcfadden1990stochastic}'s Axiom of Revealed Stochastic Preference; each of these results straightforwardly yield analogous results about SU over two periods. I characterize SU without cardinality restrictions over arbitrary finite time periods using an analog of \cite{clark1996random}'s Coherency axiom. Finally, \cite{chambers2021correlated} also study correlated random choice over lotteries over finite prize sets and characterize a model of correlated random expected utility.

The rest of the paper proceeds as follows. \textbf{Section \ref{sec:dru}} defines the choice environment, primitive, and model. \textbf{Section \ref{sec:results}} states the axioms and main results. \textbf{Section \ref{sec:app}} contains some useful auxiliary results, including the joint, marginal, and conditional Block--Marschak identities, and all proofs.
\section{Stochastic Utility}\label{sec:dru}
\subsection{Primitive}
There are $n$ time periods, indexed by $t=1,\ldots,n$. For each $t$, let $X_t$ be a finite choice set, $\mathcal{M}_t$ be the set of nonempty subsets of $X_t$, and $\Delta(X_t)$ be the set of probability distributions over $X_t$. Given a finite choice set $X$ and the set of its nonempty subsets $\mathcal{M}$, say that $\rho: \mathcal{M} \rightarrow X$ is a (static) \textit{stochastic choice function (SCF)} if $\rho(\cdot,A) \in \Delta(A)$ for each $A \in \mathcal{M}$. 

\begin{definition}
A \textbf{dynamic stochastic choice function (DSCF)} is a tuple $\rho:=(\rho_1,\ldots,\rho_n)$ such that $\rho_1: \mathcal{M}_1 \rightarrow \Delta(X_1)$ is a SCF and, for each $1<t\leq n$, $\rho_t: \mathcal{H}_{t-1} \times \mathcal{M}_t\rightarrow \Delta(X_t)$ maps each $t-1$-\textbf{history} to the SCF $\rho_t(\cdot|h_{t-1})$,\footnote{Note that the domain of $\rho_t(\cdot|h_{t-1})$ is the full set of period-$t$ menus $\mathcal{M}_t$, independently of which history $h_{t-1}$ is observed. In \cite{frick2019dynamic}'s setup, the domain of $\rho_t(\cdot|h_{t-1})$ is itself a function of $h_{t-1}$ and need not be the entire set $\mathcal{M}_t$. These features of their DSCF illustrate endogenous menu variation and limited observability, respectively.} where the set of $t-1$-histories is defined recursively:\footnote{Let $\mathcal{H}_1:=\{(A_1,x_1) \in \mathcal{M}_1\times X_1: \rho_1(x_1,A_1)>0\}$.}
$$\mathcal{H}_{t-1}:=\{(A_{t-1},x_{t-1};h_{t-2}) \in \mathcal{M}_{t-1}\times X_{t-1}\times \mathcal{H}_{t-2}: \rho_{t-1}(x_{t-1},A_{t-1}|h_{t-2})>0\}$$
\end{definition}

For each $1 \leq t \leq n$, let poset $(L_t,\leq_t):=(2^{X_t},\subseteq)$. Define their product poset to be $(L,\leq)$, where $L=\times_{t=1}^n L_t$ and $A \leq B$ if and only if $A_t \subseteq B_t$ for each $1 \leq t \leq n$. Given a set of times $T\subseteq N:=\{1,\ldots,n\}$, let subscript $T$ denote a vector indexed by $t\in T$, and let $-t:=N\backslash\{t\}$. For vectors indexed by $N$, I omit the subscript entirely. Say that $A_T<<B_T$ if $A_t\subsetneq B_t$ for all $t \in T$, and say that $y_T\neq\neq x_T \in X_T$ if $y_t\neq x_t$ for all $t \in T$.

A \textit{choice path} is a tuple $(A,x)^t:=(A_t,x_t;\ldots;A_1,x_1)$. Let $(A,x)$ denote an $n$-period choice path. A \textit{zero-probability choice path} is a choice path $(A,x)^t$ such that $\rho_1(x_1,A_1)=0$ or $\rho_s(x_s,A_s|(A,x)^{s-1})=0$ for some $1<s \leq t$. For each $1<t\leq n$, each zero-probability choice path $(A,x)^{t-1}$, and each $A_t \in \mathcal{M}_t$, define $\rho_t(\cdot,A_t|(A,x)^{t-1})$ to be any probability distribution over $A_t$. I take this augmented DSCF $\rho$ as primitive. Note that for any $A_{\leq t}:=(A_1,\ldots,A_t)$, $\rho$ admits a well-defined joint distribution $p_t(\cdot|A_{\leq t}) \in \Delta(A_{\leq t})$, defined as $p_t(x_{\leq t},A_{\leq t}):=\rho_1(x_1,A_1)\prod_{s=2}^t \rho_s(x_s,A_s|(A,x)^{s-1})$. Let $p:=p_n$. Note that for any $1 \leq r<t\leq n$ and any $(A_1,\ldots,A_t)$, $p_r$ is the marginal distribution of $p_t$ on $\times_{s=1}^r A_s$:
\begin{align*}
    \sum_{x_{>r} \in A_{>r}} p_t(x_{\leq r},A_{\leq r};x_{>r},A_{>r})=\rho_1(x_1,A_1)\prod_{s=2}^r \rho_s(x_s,A_s|(A,x)^{s-1}) \ \times \\
    \sum_{x_{r+1} \in A_{r+1}} \bigg[\rho_{r+1}(x_{r+1},A_{r+1}|(A,x)^r)\bigg[\cdots\bigg[\sum_{x_t \in A_t} \rho_t(x_t,A_t|(A,x)^{t-1}) \bigg]\cdots\bigg]\bigg] \\
    =\rho_1(x_1,A_1)\prod_{s=2}^r \rho_s(x_s,A_s|(A,x)^{s-1})=p_r(x_{\leq r},A_{\leq r})
\end{align*}

\begin{definition}
Given $A<<X$ and $x \in A^C$, their \textbf{joint Block--Marschak (BM) sum} is
$$m(x,A):=\sum_{B\geq A^C} (-1)^{\sum_{t=1}^n (|B_t|-|A_t^C|)} p(x,B)$$
\end{definition}

Joint BM sums are the multi-period analog of \cite{block1959random}'s BM sums. As in the static case, many of my forthcoming axioms will impose discipline on joint BM sums. Unlike the static case, these axioms alone are insufficient for SU.

\subsection{Model}
For each $1 \leq t \leq n$, let $P_t$ be the set of strict preference relations on $X_t$ and let $P_T:=\times_{t\in T} P_t$. Given $x_t \notin A_t \in \mathcal{M}_t$, say that $x_t \succ_t A_t$ if $x_t\succ_t y_t$ for all $y_t \in A_t$. Given $x_t \in A_t \in \mathcal{M}_t$, let $C_t(x_t,A_t):=\{\succ_t \ \in P_t: x_t \succ_t A_t\backslash\{x_t\}\}$, and let $C(x_t,A_t):=C_t(x_t,A_t)\times P_{-t}$. Given history $h_t:=(A,x)^t$, let $C(h_t):=\bigcap_{s=1}^t C(x_s,A_s)$. Given $x_T \in A_T$, let $C(x,A)^T:=\bigcap_{t\in T} C(x_t,A_t)=\times_{t\in T} C_t(x_t,A_t) \times P_{-T}$. Given $T' \subseteq T\subseteq N$, let $C_T(x_{T'},A_{T'}):=\times_{t \in T'}C_t(x_t,A_t)\times P_{T\backslash T'}$.

\begin{definition}
A \textbf{stochastic utility (SU) representation} of $\rho$ is a probability measure $\mu \in \Delta(P)$ such that $\rho_1(x_1,A_1)=\mu(C(x_1,A_1))$ for all $x_1 \in A_1 \in \mathcal{M}_1$ and
$$\rho_t(x_t,A_t|h_{t-1})=\mu(C(x_t,A_t)|C(h_{t-1}))$$
for all $1<t\leq n$, $h_{t-1} \in \mathcal{H}_{t-1}$, and $x_t \in A_t \in \mathcal{M}_t$.
\end{definition}

Given $A_t\subsetneq X_t$ and $x_t \in A_t^c$, define their \textit{joint upper edge set} to be $E_t(x_t,A_t):=\{\succ_t \in P_t: A_t \succ_t x_t \succ_t A_t^C\backslash\{x_t\}\}$. In words, this is the set of period-$t$ preferences that rank $x_t$ on the uppermost edge of $A_t^C$ and below the lowermost edge of $A$. Given $A_T<<X_T$ and $x_T \in A_T^C$, define $E(x_t,A_t):=E_t(x_t,A_t)\times P_{-t}$ and $E(x,A)^T=\bigcap_{t \in T} E(x_t,A_t)$. Given $T' \subseteq T\subseteq N$, let $E_T(x_{T'},A_{T'}):=\times_{t \in T'}E_t(x_t,A_t) \times P_{T\backslash T'}$.
\section{Results}\label{sec:results}
The first result characterizes SU representations as probability measures that assign every joint upper edge set its corresponding joint Block--Marschak sum. It also establishes an equivalence between SU representations of $\rho$ and CRUM representations of the induced correlated choice rule $p$, as defined in \cite{chambers2021correlated}.

\begin{proposition}\label{p:SU_iff_assigns}
The following are equivalent.
\begin{enumerate}
    \item $\mu$ is an SU representation of $\rho$.
    \item $\mu(C(x,A))=p(x,A)$ for all $x \in A \in \mathcal{M}$.
    \item $\mu(E(x,A))=m(x,A)$ for all $A<<X$ and $x \in A^C$.
\end{enumerate}
\end{proposition}

The next result uses the following axiom to characterize SU when all choice sets have three elements. I abuse notation to let $\{x\}^C=(\{x_t\}^C)_{t=1}^n$.

\begin{axiom}[\textbf{Joint Supermodularity}]
For all $1 \leq t \leq n$ and $y_t \neq x_t \in X_t$,
$$\sum_{B\geq \{x\}^C: \sum_{t=1}^n |B_t| \text{even}} p(y,B) \geq \sum_{B\geq \{x\}^C: \sum_{t=1}^n |B_t| \text{odd}} p(y,B)$$
\end{axiom}

\begin{example}
\normalfont Let $n=2$ and $|X_1|=|X_2|=3$. Joint Supermodularity requires that for any $b \neq a \in X_1$ and $y \neq x \in X_2$,
\begin{align*}
    p(b,\{b,c\};y,\{y,z\})+p(b,X_1;y,X_2) \geq p(b,\{b,c\};y,X_2)+p(b,X_1;y,\{y,z\})
\end{align*}
\end{example}

\begin{theorem}\label{p:all3}
Suppose $|X_t|=3$ for all $1 \leq t \leq n$. $\rho$ has a unique SU representation if and only if it satisfies Joint Supermodularity and Marginal Consistency.
\end{theorem}

The proof of this result proceeds from the observation that when all choice sets have three elements, the candidate SU representation is pinned down by joint BM sums of the form $m(y,\{x\})$, where I again abuse notation and let $\{x\}=(\{x_t\})_{t=1}^n$. The next result characterizes SU under a relaxation of the last period's cardinality restrictions with the following two axioms.

\begin{axiom}[\textbf{Joint BM Nonnegativity}]
$m(x,A)\geq0$ for all $A<<X$ and $x \in A^C$.
\end{axiom}

If $m(x,A)>0$ for all $A<<X$ and $x \in A^C$, say that $\rho$ satisfies \textit{Joint BM Positivity}. The two-period version of this axiom is equivalent to \cite{chambers2021correlated}'s Axiom 3. Note that when all choice sets have three elements, Joint BM Nonnegativity implies Joint Supermodularity, since $\sum_{B\geq \{x\}^C: \sum_{t=1}^n |B_t| \text{even}} p(y,B)-\sum_{B\geq \{x\}^C: \sum_{t=1}^n |B_t| \text{odd}} p(y,B)=\sum_{B\geq\{x\}^C} (-1)^{\sum_{t=1}^n |B_t|-2n}p(y,B)=m(y,\{x\})\geq 0$. As I show in the proof of \textbf{Theorem \ref{p:all3}}, Joint Supermodularity and the following axiom imply Joint BM Nonnegativity under these cardinality restrictions.

\begin{axiom}[\textbf{Marginal Consistency}]\label{ax:mc}
For all $1 \leq t<n$ and $x_{-t} \in A_{-t} \in \mathcal{M}_{-t}$,
$$\sum_{x_t \in A_t} p(x_t,A_t;x_{-t},A_{-t})$$
is constant in $A_t$.\footnote{For $x_{-n} \in A_{-n} \in \mathcal{M}_{-n}$ and any $A_n$, we have $\sum_{x_n \in A_n} p(x_{-n},A_{-n};x_n,A_n)=p_{n-1}(x_{-n},A_{-n})$, since $p_{n-1}$ is the marginal of $p$ on $A_{-n}$.}
\end{axiom}

If $p$ satisfies Marginal Consistency, define $p(x_{-t},A_{-t}):=p(x_t,\{x_t\};x_{-t},A_{-t})$ for any $x_t \in X_t$. For any $T\subsetneq \{1,\ldots,n\}$, $A_T=(A_t)_{t \in T}$, and $A_{-T}=(A_t)_{t \notin T}$, I show via induction on the size of $T$ that Marginal Consistency implies a well-defined marginal distribution over $A_{-T}$, defined by $p(x_{-T},A_{-T};y_T,\{y\}_T)$ for any $y_T \in X_T$. The base case ($|T|=1$) follows directly from the content of Marginal Consistency. Inductive step ($1<|T|<n$): fix $t \in T$, $y_t \in X_t$, and $y_{T\backslash\{t\}} \in X_{T\backslash\{t\}}$ Then
\begin{align*}
    \sum_{x_T \in A_T} p(x_T,A_T;x_{-T},A_{-T})=\sum_{x_t \in A_t} \sum_{x_{T\backslash\{t\}} \in A_{T\backslash\{t\}}} p(x_t,A_t;x_{T\backslash\{t\}},A_{T\backslash\{t\}};x_{-T},A_{-T}) \\
    =\sum_{x_t \in A_t} p(x_t,A_t;y_{T\backslash\{t\}},\{y\}_{T\backslash\{t\}};x_{-T},A_{-T})=p(y_T,\{y\}_T;x_{-T},A_{-T})
\end{align*}
where the second equality follows from the inductive hypothesis and the third equality follows from Marginal Consistency. To compensate for the latent last-period marginality I assume in $p$, my axiom of Marginal Consistency is weaker than \cite{chambers2021correlated}'s Axiom 2 of Marginality.

\begin{theorem}\label{thm}
Suppose $|X_t|=3$ for all $1\leq t<n$. $\rho$ has an SU representation if and only if it satisfies Joint BM Nonnegativity and Marginal Consistency.\footnote{A previous draft of this paper incorrectly stated this result without the cardinality restrictions.}
\end{theorem}

The proof of the backwards direction of \textbf{Theorem \ref{thm}} proceeds through a series of five claims. First, I recursively define a set function $\nu$, whose domain is not the entire power set $2^P$ but instead tuples of \textit{cylinders}. The first two claims enforce two necessary additive properties of $\nu$. The next two claims allow for the construction of a probability measure $\mu \in \Delta(P)$ that extends $\nu$. The last claim verifies that $\mu$ assigns each joint upper edge set its corresponding joint BM sum. By sharpening Joint BM Nonnegativity to Positivity, \textbf{Theorem \ref{thm}} admits a corollary that characterizes SU with full support.

\begin{corollary}\label{c:fullsup}
Suppose $|X_t|=3$ for all $1\leq t<n$. $\rho$ has an SU representation with full support over $P$ if and only if it satisfies Joint BM Positivity and Marginal Consistency.
\end{corollary}

Next, I present an alternative characterization of SU under the same cardinality restrictions. For any $A_{-n}<<X_{-n}$ and $x_{-n} \in A_{-n}^C$, define their \textit{marginal Block--Marschak (BM) sum} to be
\begin{align*}
    m(x_{-n},A_{-n}):=\sum_{x_n \in X_n} m(x_{-n},A_{-n};x_n,\emptyset)=\sum_{B_{-n} \geq A_{-n}^C} (-1)^{\sum_{t=1}^{n-1} |B_t|-|A_t^C|} p_{-n}(x_{-n},B_{-n})
\end{align*}

Recall that $p_{-n}:=p_{n-1}$ is well-defined without assuming Marginal Consistency of $p$.

\begin{axiom}[Partial Marginal BM Nonnegativity]\label{ax:pmbmn}
For all $1\leq t<n$ and $y_t\neq x_t \in X_t$, $m(y_{-n},\{x\}_{-n})\geq 0$.
\end{axiom}

Note that Joint BM Nonnegativity implies Partial Marginal BM Nonnegativity, since each marginal BM sum is itself a sum of joint BM sums. Next, for any $A_{-n}<<X_{-n}$ and $x_{-n} \in A_{-n}^C$ with $m(x_{-n},A_{-n})>0$, define their \textit{conditional SCF}\footnote{Strictly speaking, $\rho_n(x_n,A_n|(x,A)^{-n})$ need not be nonnegative, even assuming Partial Marginal BM Nonnegativity. Let $n=2$, $|X_1|=3$, $(x,A)^{-n}=(y_1,\{x_1\})$, and $x_2 \in A_2=X_2$: then $\rho_2(x_2,A_2|(y_1,\{x_1\}))=\frac{p(y_1,z_1;x_2,A_2)-p(y_1,X_1;x_2,A_2)}{\rho_1(y_1,z_1)-\rho_1(y_1,X_1)}$. Let $\rho_1(y_1,z_1)=\frac{1}{2}$, $\rho_2(x_2,A_2|\{y_1,z_1\},y_1)=0$, $\rho_1(y_1,X_1)=\frac{1}{4}$, and $\rho_2(x_2,A_2|X_1,y_1)=1$. The following axiom enforces that conditional SCFs are indeed SCFs.} as
\begin{align*}
    \rho_n(x_n,A_n|(x,A)^{-n}):=\frac{\sum_{D_n\subseteq A_n^C} m(x_{-n},A_{-n};x_n,D_n)}{m(x_{-n},A_{-n})}
\end{align*}
for any $x_n \in A_n \in \mathcal{M}_n$\footnote{If $x_n\notin A_n$, define $\rho_n(x_n,A_n|(x,A)^{-n}):=0$.}
and their \textit{conditional Block--Marschak sum} as 
\begin{align*}
    m(x_n,A_n|(x,A)^{-n}):=\sum_{B_n \supseteq A_n^C} (-1)^{|B_n|-|A_n^C|} \rho_n(x_n,B_n|(x,A)^{-n})
\end{align*}
for any $A_n\subsetneq X_n$ and $x_n \in A_n^C$.

\begin{axiom}[Partial Conditional BM Nonnegativity]\label{ax:pcbmn}
For all $1\leq t<n$, $y_t\neq x_t \in X_t$ with $m(y_{-n},\{x\}_{-n})>0$, $A_n\subsetneq X_n$, and $x_n\in A_n^C$, $m(x_n,A_n|(y,\{x\})^{-n})\geq0$.
\end{axiom}

\textbf{Lemma \ref{l:condSCFBM}} shows that $m(x_n,A_n|(x,A)^{-n})=\frac{m(x,A)}{m(x_{-n},A_{-n})}$ and hence Joint BM Nonnegativity implies Partial Conditional BM Nonnegativity.

\begin{axiom}[$(-n)$-Marginal Consistency]\label{ax:-nmc}
$p_{-n}$ satisfies Marginal Consistency.
\end{axiom}

By the argument following \textbf{Axiom \ref{ax:mc}}, Marginal Consistency of $p$ implies $(-n)$-Marginal Consistency.\footnote{In the context of that argument, take $T=\{t,n\}$ for any $1\leq t<n-1$. To see that the converse does not hold in general, let $n=2$: then $(-n)$-Marginal Consistency is satisfied by definition of $\rho_1$, but $\sum_{x_1 \in A_1} p(x_1,A_1;x_2,A_2)$ need not be constant in $A_1$.} Finally, say that $(x_{-n}^i,A_{-n}^i)_{i \in I}$ \textit{partition} history $h_{n-1} \in \mathcal{H}_{n-1}$ if $\{E(x_{-n}^i,A_{-n}^i)\}$ is a partition of $C(h_{n-1})$. Given partition $(x_{-n}^i,A_{-n}^i)_{i \in I}$ of $h_{n-1}$, let $I'=\{i \in I: m(x_{-n}^i,A_{-n}^i)>0\}$.

\begin{axiom}[$(-n)$-Conditional Consistency]\label{ax:-ncondcon}
For any $x_n \in A_n \in \mathcal{M}_n$ and $(A,x)^{-n} \in \mathcal{H}_{n-1}$ with partition $(y_{-n}^i,\{x^i\}_{-n})_{i \in I}$, 
\begin{align*}
    \rho_n(x_n,A_n|(A,x)^{-n})=\sum_{i \in I'} \rho_n(x_n,A_n|(y^i,\{x^i\})^{-n})\frac{m(y_{-n}^i,\{x^i\}_{-n})}{p_n(x_{-n},A_{-n})}
\end{align*}
\end{axiom}

This axiom is similar in spirit to \cite{frick2019dynamic}'s Linear History Independence axiom. However, since my choice environment consists of riskless finite sets, the set of observable histories is coarser than that of \cite{frick2019dynamic}'s setup. In particular, $\rho_n(\cdot|(y^i,\{x^i\})^{-n})$ must be counterfactually extrapolated, whereas the analog of this SCF in \cite{frick2019dynamic}'s environment can be directly observed.

\begin{theorem}\label{SU4axioms}
Suppose $|X_t|=3$ for all $1\leq t<n$. $\rho$ has an SU representation if and only if it satisfies \textbf{Axioms \ref{ax:pmbmn}-\ref{ax:-ncondcon}}.
\end{theorem}

The proof of the backwards direction of \textbf{Theorem \ref{SU4axioms}} proceeds by constructing a marginal SU representation $\mu_{-n} \in \Delta(P_{-n})$ for the first $n-1$ periods, a conditional RU representation $\mu^{\succ_{-n}} \in \Delta(P_n)$ for each $\succ_{-n} \in \text{supp } \mu_{-n}$, and finally a candidate SU representation $\mu \in \Delta(P)$ that mixes the conditional RU representations $(\mu^{\succ_{-n}})$ according to $\mu_{-n}$. Finally, I characterize SU for arbitrary finite choice sets over an arbitrary finite time horizon, using the following axiom.

\begin{axiom}[Joint Coherency]
For any $(x^i,A^i)_{i=1}^k$ with $x^i \in A^i \in \mathcal{M}$ for each $1\leq i\leq k$ and for any $(\lambda^i)_{i=1}^k \subseteq \R$,
\begin{align*}
    \sum_{i=1}^k \lambda^i\mathbbm{1}_{C(x^i,A^i)} \geq 0 \implies \sum_{i=1}^k \lambda^i p(x^i,A^i) \geq 0
\end{align*}
\end{axiom}

Joint Coherency is the multiperiod analog of \cite{clark1996random}'s Coherency axiom, which in turn is based on \cite{de1937prevision}'s characterization of finitely additive probability measures on an algebra of sets.

\begin{theorem}\label{t:jcoh}
$\rho$ has an SU representation if and only if it satisfies Joint Coherency.
\end{theorem}
\section{Appendix}\label{sec:app}

\subsection{The Möbius Inversion}
Let $(L,\leq)$ be a finite, partially ordered set (poset). The following definition and lemma are due to \cite{van2001course} Equations 25.2 and 25.5.

\begin{definition}
The \textbf{Möbius function} $m_L: L^2 \rightarrow \mathbb{Z}$ is
\begin{align*}
    m_L(a,b)=\begin{cases} 
      1 & a=b \\
      0 & a \nleq b \\
      -\sum_{a \leq c<b} m_L(a,c) & a<b
   \end{cases}
\end{align*}
\end{definition}

Let the \textit{zeta function} of $L$, denoted $\zeta_L$, be the indicator function for the set $\leq \ \subseteq L^2$. By \cite{van2001course} Equation 25.1, $m_L$ is the $|L| \times |L|$ matrix inverse of $\zeta_L$.

\begin{lemma}\label{l:inv}
Given a function $f:L \rightarrow \mathbb{R}$, define $F(a):=\sum_{b \geq a} f(b)$. Then
\begin{align*}
    f(a)=\sum_{b \geq a} m_L(a,b)F(b)
\end{align*}
\end{lemma}

\begin{proof}
This proof is adapted from page 336 of \cite{van2001course}. Fix $f: L \rightarrow \R$ and define $F$ as above. Observe that
\begin{align*}
    \sum_{a \leq c \leq b} m_L(a,c)=\sum_{c \in L} m_L(a,c)\zeta_L(c,b)=\begin{cases} 
      1 & a=b \\
      0 & \text{else}
   \end{cases}
\end{align*}
where the first equality holds because $m_L(a,c)=0$ if $a \nleq c$ and $\zeta_L(c,b)=1$ if $c\leq b$ and $0$ else, and the second equality holds because $m_L\zeta_L$ is the $L \times L$ identity matrix. Thus,
\begin{align*}
    \sum_{b \geq a} m_L(a,b)F(b)=\sum_{b \geq a} m_L(a,b)\bigg(\sum_{c \geq b} f(c)\bigg)=\sum_{c\geq a} f(c) \bigg(\sum_{a \leq b \leq c} m_L(a,b)\bigg)=f(a)
\end{align*}
\end{proof}

\textbf{Lemma \ref{l:inv}} shows how to recover any real-valued poset-defined function given its sums over upper contour sets and the poset's Möbius function. This procedure is known as the \textit{Möbius inversion}. The following lemma is due to \cite{van2001course} Theorem 25.1(i). It states the Möbius function for the power set of a finite set under the inclusion partial order.

\begin{lemma}\label{l:power}
Fix finite $X$ and let $L=2^X$ with $\leq=\subseteq$. Then
\begin{align*}
    m_L(A,B)=\begin{cases} 
      (-1)^{|B|-|A|} & A \subseteq B \\
      0 & \text{else}
   \end{cases}
\end{align*}
\end{lemma}

\begin{proof}
See page 343 of \cite{van2001course}.
\end{proof}

Given posets $(L_i,\leq_i)_{i=1}^n$, define their \textit{product poset} to be $(L,\leq)$, where $L=\times_{i=1}^n L_i$ and $a \leq b$ if and only if $a_i \leq_i b_i$ for each $i$. Let $m_i$ denote the Möbius function of $L_i$. The following lemma generalizes \cite{godsil2018introduction} Lemma 3.1 to state the Möbius function for $n$-ary product posets for arbitrary finite $n$.

\begin{lemma}\label{l:prod}
For all $a,b \in L$, the Möbius function $m_L: L^2 \rightarrow \mbb{Z}$ satisfies
\begin{align*}
    m_L(a,b)=\prod_{i=1}^n m_{L_i}(a_i,b_i)
\end{align*}
\end{lemma}

\begin{proof}
Let $a,b \in L$. By definition of $\leq_L$, $\zeta_L(a,b)=1$ if and only if $\zeta_{L_i}(a_i,b_i)=1$ for each $i$. Hence, $\zeta_L(a,b)=\prod_{i=1}^n \zeta_{L_i}(a_i,b_i)$, which implies the $|L|\times|L|$ matrix $\zeta_L$ is the \textit{Kronecker product} of the $|L_i|\times |L_i|$ matrices $\zeta_{L_i}$, denoted $\bigotimes_{i=1}^n \zeta_{L_i}$.\footnote{See Definition 2.1 of \cite{schacke2004kronecker} for a definition of the Kronecker product.} Then 
$$m_L=(\zeta_L)^{-1}=\bigg(\bigotimes_{i=1}^n \zeta_{L_i}\bigg)^{-1}=\bigotimes_{i=1}^n (\zeta_{L_i})^{-1}=\bigotimes_{i=1}^n m_{L_i}$$
where the third equality follows from KRON 10 of \cite{schacke2004kronecker}.\footnote{Formally, I can show this by inducting on $n$. The base case ($n=1$) immediately follows. Inductive step: $(\bigotimes_{i=1}^n \zeta_{L_i})^{-1}=((\bigotimes_{i=1}^{n-1} \zeta_{L_i}) \otimes \zeta_{L_n})^{-1}=(\bigotimes_{i=1}^{n-1} \zeta_{L_i})^{-1} \otimes (\zeta_{L_n})^{-1}=\bigotimes_{i=1}^n (\zeta_{L_i})^{-1}$, where the second equality follows from KRON 10 of \cite{schacke2004kronecker} and the third equality follows from the inductive hypothesis.} By definition of the Kronecker product, we conclude
\begin{align*}
    m_L(a,b)=\prod_{i=1}^n m_{L_i}(a_i,b_i)
\end{align*}.
\end{proof}

\subsection{BM Sum Identities}
The following lemma shows how to recover $p$ from lower contour sums of joint BM sums.

\begin{lemma}\label{p_BM_sums}
For all $A<<X$ and $x \in A^C$,
\begin{align*}
    p(x,A^C)=\sum_{B \leq A} m(x,B)
\end{align*}
\end{lemma}

\begin{proof}
By \textbf{Lemmas \ref{l:power} and \ref{l:prod}}, I obtain
\begin{align*}
    m_L(A,B)=\begin{cases} 
      (-1)^{\sum_{t=1}^n (|B_t|-|A_t|)} & A \leq B \\
      0 & \text{else}
   \end{cases}
\end{align*}
For each $1 \leq t \leq n$, fix any $A_t \subsetneq X_t$ and $x_t \in A_t^C$. Define $f: L \rightarrow \mathbb{R}$ as
\begin{align*}
    f(B):=(-1)^{\sum_{t=1}^n (|B_t|-|A_t^C|)} p(x,B)
\end{align*}
and $F: L \rightarrow \mathbb{R}$ as $F(D):=\sum_{B \geq D} f(B)$. Applying the Möbius inversion from \textbf{Lemma \ref{l:inv}},
\begin{align*}
    f(D)=\sum_{B \geq D} (-1)^{\sum_{t=1}^n (|B_t|-|D_t|)} F(B)
\end{align*}
which implies
\begin{align*}
    p(x,A^C)=f(A^C)
    =\sum_{B \geq A^C} (-1)^{\sum_{t=1}^n (|B_t|-|A_t^C|)} F(B)=\sum_{D \leq A} m(x,D)
\end{align*}
The last equality follows by matching terms across sums via the bijection $D^C \geq A^C \iff D \leq A$:
\begin{align*}
    (-1)^{\sum_{t=1}^n (|D_t^C|-|A_t^C|)} F(D^C)=\sum_{B \geq D^C} (-1)^{\sum_{t=1}^n (|D_t^C|-|A_t^C|)} f(B) \\
    =\sum_{B\geq D^C} (-1)^{\sum_{t=1}^n (|B_t|-|D_t^C|)} p(x,B)=m(x,D)
\end{align*}
where the second equality follows by observing that $(-1)^k=(-1)^{-k}$.
\end{proof}

\textbf{Lemma \ref{p_BM_sums}} is the dynamic analog of \cite{chambers2016revealed}'s Lemma 7.4(I), which is stated in that reference without proof. An analogous argument shows how to recover $\rho_n(\cdot|h_{t-1})$ from lower contour sums of (history-)conditional BM sums.

\begin{lemma}\label{l:1perBMsum}
For all $(A,x)^{-n} \in \mathcal{H}_{n-1}$, $A_n\subsetneq X_n$ and $x_n \in A_n^C$,
$$\rho_n(x_n,A_n^C|(A,x)^{-n})=\sum_{B_n \subseteq A_n} m(x_n,B_n|(A,x)^{-n}) $$
where
$$m(x_n,B_n|(A,x)^{-n}):=\sum_{D_n\supseteq B_n^C} (-1)^{|D_n|-|B_n^C|} \rho_n(x_n,D_n|(A,x)^{-n})$$
\end{lemma}

\begin{proof}
By \textbf{Lemma \ref{l:power}}, the Möbius function for $L_n=(2^{X_n},\subseteq)$ is
\begin{align*}
    m_{L_n}(A_n,B_n)=\begin{cases} 
      (-1)^{|B_n|-|A_n|} & A_n \subseteq B_n \\
      0 & \text{else}
   \end{cases}
\end{align*}
Fix any $(A,x)^{-n} \in \mathcal{H}_{n-1}$, $A_n\subsetneq X_n$ and $x_n \in A_n^C$. Define $f: L_n \rightarrow \mathbb{R}$ as
\begin{align*}
    f(B_n):=(-1)^{|B_n|-|A_n^C|}\rho_n(x_n,B_n|(A,x)^{-n})
\end{align*}
and $F: L_n \rightarrow \mathbb{R}$ as $F(D_n):=\sum_{B_n \supseteq D_n} f(B_n)$. Applying the Möbius inversion from \textbf{Lemma \ref{l:inv}},
\begin{align*}
    f(D_n)=\sum_{B_n \supseteq D_n} (-1)^{|B_n|-|D_n|} F(B_n)
\end{align*}
which implies
\begin{align*}
    \rho_n(x_n,A_n^C|(A,x)^{-n})=f(A_n^C)
    =\sum_{B_n \supseteq A_n^C} (-1)^{|B_n|-|A_n^C|} F(B_n)=\sum_{D_n \subseteq A_n} m(x_n,D_n|(A,x)^{-n})
\end{align*}
The last equality follows by matching terms across sums via the bijection $D_n^C \supseteq A_n^C \iff D_n \subseteq A_n$:
\begin{align*}
    (-1)^{|D_n^C|-|A_n^C|} F(D_n^C)=\sum_{B_n \supseteq D_n^C} (-1)^{|D_n^C|-|A_n^C|} f(B_n) \\
    =\sum_{B_n\supseteq D_n^C} (-1)^{|B_n|-|D_n^C|} \rho_n(x_n,B_n|(A,x)^{-n})=m(x_n,D_n|(A,x)^{-n})
\end{align*}
where the second equality follows by observing that $(-1)^k=(-1)^{-k}$.
\end{proof}

Strictly speaking, \textbf{Lemma \ref{l:1perBMsum}} holds for SCFs and BM sums that are conditional on histories. However, the same argument implies that for any $A<<X$ and $x\notin A$ with $m(x_{-n},A_{-n})>0$,
$$\rho_n(x_n,A_n^C|(x,A)^{-n})=\sum_{B_n\subseteq A_n} m(x_n,B_n|(x,A)^{-n})$$
This follows by fixing $A<<X$ and $x\notin A$ with $m(x_{-n},A_{-n})>0$, defining $f(B_n):=(-1)^{|B_n|-|A_n^C|}\rho_n(x_n,B_n|(x,A)^{-n})$ and $F(D_n):=\sum_{B_n\supseteq D_n} f(B_n)$, applying the Möbius inversion as before, and concluding that $\rho_n(x_n,A_n^C|(x,A)^{-n})=\sum_{B_n \supseteq A_n^C} (-1)^{|B_n|-|A_n^C|} F(B_n)=\sum_{D_n \subseteq A_n} m(x_n,D_n|(x,A)^{-n})$ via matching terms: for any $D_n^C\supseteq A_n^C$,
\begin{align*}
    (-1)^{|D_n^C|-|A_n^C|} F(D_n^C)=\sum_{B_n \supseteq D_n^C} (-1)^{|D_n^C|-|A_n^C|} f(B_n) \\
    =\sum_{B_n\supseteq D_n^C} (-1)^{|B_n|-|D_n^C|} \rho_n(x_n,B_n|(x,A)^{-n})=m(x_n,D_n|(x,A)^{-n})
\end{align*}

The following lemma equates two different (single-period) sums of joint BM sums under Marginal Consistency.

\begin{lemma}\label{l:sums_BM_sums}
Assume $p$ satisfies Marginal Consistency. For any $1 \leq t \leq n$, $\emptyset \subsetneq A_t \subsetneq X_t$, $A_{-t}<<X_{-t}$, and $x_{-t} \notin A_{-t}$,
$$\sum_{x_t \in A_t^C} m(x_t,A_t;x_{-t},A_{-t})=\sum_{y_t \in A_t} m(y_t,A_t\backslash\{y_t\};x_{-t},A_{-t})$$
\end{lemma}

\begin{proof}
Expanding the left-hand side yields
$$\sum_{B_{-t}\geq A_{-t}^C} (-1)^{\sum_{i \neq t} |B_i|-|A_i^C|} \bigg[\sum_{B_t \supseteq A_t^C} (-1)^{|B_t|-|A_t^C|}p(A_t^C,B_t;x_{-t},B_{-t})\bigg]$$
and expanding the right-hand side yields
$$\sum_{B_{-t}\geq A_{-t}^C} (-1)^{\sum_{i \neq t} |B_i|-|A_i^C|} \bigg[\sum_{y_t \in A_t} \sum_{B_t \supseteq A_t^C\cup\{y_t\}} (-1)^{|B_t|-|A_t^C|-1} p(y_t,B_t;x_{-t},B_{-t})\bigg]$$

I will show that the inner sums are equal by matching terms. Expanding the first inner sum yields
\begin{align*}
    p(x_{-t},B_{-t})+\sum_{B_t\supsetneq A_t^C} (-1)^{|B_t|-|A_t^C|}\big[p(x_{-t},B_{-t})-p(B_t\backslash A_t^C,B_t;x_{-t},B_{-t})\big] \\
    =\sum_{B_t \supseteq A_t^C} (-1)^{|B_t|-|A_t^C|} p(x_{-t},B_{-t})-\sum_{B_t \supsetneq A_t^C} (-1)^{|B_t|-|A_t^C|} p(B_t\backslash A_t^C,B_t;x_{-t},B_{-t}) \\
    =p(x_{-t},B_{-t})\sum_{B_t \supseteq A_t^C} (-1)^{|B_t|-|A_t^C|}+\sum_{B_t \supsetneq A_t^C} (-1)^{|B_t|-|A_t^C|+1} p(B_t\backslash A_t^C,B_t;x_{-t},B_{-t}) \\
    =\sum_{B_t \supsetneq A_t^C} \sum_{x_t \in B_t\backslash A_t^C} (-1)^{|B_t|-|A_t^C|+1} p(x_t,B_t;x_{-t},B_{-t})
\end{align*}
where the first line follows from Marginal Consistency. To see the last equality, observe that
$$\sum_{B_t \supseteq A_t^C} (-1)^{|B_t|-|A_t^C|}=\sum_{k=0}^{|A_t|} (-1)^k \binom{|A_t|}{k}=0$$
since, for each $0\leq k\leq |A_t|$, there are $\binom{|A_t|}{k}$ supersets of $A_t^C$ with $|A_t^C|+k$ elements. Since $(-1)^n=(-1)^{n+2}$, matching terms across inner sums via the bijection $y_t \in A_t, B_t \supseteq A_t^C \cup \{y_t\} \iff B_t \supsetneq A_t^C, y_t \in B_t\backslash A_t^C$ completes the proof.
\end{proof}

\textbf{Lemma \ref{l:sums_BM_sums}} is the dynamic analog of \cite{chambers2016revealed}'s Lemma 7.4(II), which is stated in that reference without proof. Note that since $p_{n-1}$ is the marginal of $p$ over $A_{-n}$, the result of \textbf{Lemma \ref{l:sums_BM_sums}} for $t=n$ goes through without assuming Marginal Consistency. \textbf{Lemma \ref{l:sums_BM_sums}} admits the following corollary, which equates multi-period sums of Block-Marschak sums.

\begin{corollary}\label{c:sums_BM}
Assume $p$ satisfies Marginal Consistency. For any (nonempty) set of indices $T \subseteq \{1,\ldots,n\}$, any $\emptyset<<A_T<<X_T$, $A_{-T}<<X_{-T}$, and $x_{-T}\notin\notin A_{-T}$,\footnote{Let $A_T:=(A_t)_{t \in T}$ and say that $x_T \notin\notin A_{-T}$ if $x_t \notin A_t$ for all $t \in T$.}
$$\sum_{x_T \in A_T^C} m(x_T,A_T;x_{-T},A_{-T})=\sum_{y_T \in A_T} m(y_T,(A\backslash\{y\})_T;x_{-T},A_{-T})$$
where $(A\backslash\{y\})_T:=(A_t\backslash\{y_t\})_{t \in T}$.
\end{corollary}

\begin{proof}
I prove this by inducting on the cardinality of $T$. The base case ($|T|=1$) is \textbf{Lemma \ref{l:sums_BM_sums}}. Inductive step: suppose the equality holds for all $T \subseteq \{1,\ldots,n\}$ with $|T|=k$. Fix $T=\{t_1,\ldots,t_{k+1}\} \subseteq \{1,\ldots,n\}$ and let $T_k=T\backslash\{t_{k+1}\}$. Then
\begin{align*}
    \sum_{x_T \in A_T^C} m(x_T,A_T;x_{-T},A_{-T})=\sum_{x_{t_{k+1}} \in A_{t_{k+1}}^C} \bigg[\sum_{x_{T_k} \in A_{T_k}^C} m(x_{T_k},A_{T_k};x_{t_{k+1}},A_{t_{k+1}};x_{-T},A_{-T})\bigg] \\
    =\sum_{x_{t_{k+1}} \in A_{t_{k+1}}^C} \bigg[\sum_{y_{T_k} \in A_{T_k}} m(y_{T_k},(A\backslash\{y\})_{T_k};x_{t_{k+1}},A_{t_{k+1}};x_{-T},A_{-T})\bigg] \\
    =\sum_{y_{T_k} \in A_{T_k}} \bigg[\sum_{x_{t_{k+1}} \in A_{t_{k+1}}^C} m(y_{T_k},(A\backslash\{y\})_{T_k};x_{t_{k+1}},A_{t_{k+1}};x_{-T},A_{-T})\bigg] \\
    =\sum_{y_{T_k} \in A_{T_k}} \bigg[\sum_{y_{t_{k+1}} \in A_{t_{k+1}}} m(y_{T_k},(A\backslash\{y\})_{T_k};y_{t_{k+1}},A_{t_{k+1}}\backslash\{y_{t_{k+1}}\};x_{-T},A_{-T})\bigg] \\
    =\sum_{y_T \in A_T} m(y_T,(A\backslash\{y\})_T;x_{-T},A_{-T})
\end{align*}
where the second equality follows from the inductive hypothesis and the fourth equality follows from \textbf{Lemma \ref{l:sums_BM_sums}}.
\end{proof}

The next result verifies that joint BM sums can be decomposed into products of marginal and conditional BM sums.

\begin{lemma}\label{l:condSCFBM}
For any $A_{-n}<<X_{-n}$ and $x_{-n} \in A_{-n}^C$ with $m(x_{-n},A_{-n})>0$ and any $A_n \in \mathcal{M}_n$, 
\begin{align*}
    \sum_{x_n \in A_n} \rho_n(x_n,A_n|(x,A)^{-n})=1 \ \ \ \text{and} \ \ \ m(x_n,A_n|(x,A)^{-n})m(x_{-n},A_{-n})=m(x,A)
\end{align*}
\end{lemma}

\begin{proof}
Fix any $A_{-n}<<X_{-n}$ and $x_{-n} \in A_{-n}^C$ with $m(x_{-n},A_{-n})>0$ and any $A_n \in \mathcal{M}_n$. Note that for fixed $x_n \in A_n$,
\begin{align*}
    \sum_{D_n\subseteq A_n^C} m(x_{-n},A_{-n};x_n,D_n) \\
    =\sum_{B_{-n} \geq A_{-n}^C} (-1)^{\sum_{t=1}^{n-1} |B_t|-|A_t^C|}p_{-n}(x_{-n},B_{-n})\sum_{D_n\subseteq A_n^C}\sum_{B_n\supseteq D_n^C} (-1)^{|B_n|-|D_n^C|} \rho_n(x_n,B_n|(B,x)^{-n}) \\
    =\sum_{B_{-n}\geq A_{-n}^C: p_{-n}(x_{-n},B_{-n})>0} (-1)^{\sum_{t=1}^{n-1} |B_t|-|A_t^C|} p_{-n}(x_{-n},B_{-n}) \sum_{D_n \subseteq A_n^C} m(x_n,D_n|(B,x)^{-n}) \\
    =\sum_{B_{-n}\geq A_{-n}^C} (-1)^{\sum_{t=1}^{n-1} |B_t|-|A_t^C|} p(x_{-n},B_{-n};x_n,A_n)
\end{align*}
where the second equality holds because every sum is equal to itself excluding the terms equal to zero and by the definition of $m(x_n,D_n|(B,x)^{-n})$ given in \textbf{Lemma \ref{l:1perBMsum}}, and the third equality follows from \textbf{Lemma \ref{l:1perBMsum}}. Substituting the above yields
\begin{align*}
    \sum_{x_n \in A_n} \rho_n(x_n,A_n|(x,A)^{-n})=\frac{\sum_{x_n \in A_n} \sum_{D_n\subseteq A_n^C} m(x_{-n},A_{-n};x_n,D_n)}{m(x_{-n},A_{-n})} \\
    =\frac{\sum_{B_{-n}\geq A_{-n}^C} (-1)^{\sum_{t=1}^{n-1} |B_t|-|A_t^C|} \sum_{x_n \in A_n} p(x_{-n},B_{-n};x_n,A_n)}{m(x_{-n},A_{-n})}=1
\end{align*}
by definition of $m(x_{-n},A_{-n})$. Next,
\begin{align*}
    m(x_n,A_n|(x,A)^{-n})m(x_{-n},A_{-n})=\sum_{B_n\supseteq A_n^C} (-1)^{|B_n|-|A_n^C|} \sum_{D_n\subseteq B_n^C} m(x_{-n},A_{-n};x_{n},D_n) \\
    =\sum_{B_n\supseteq A_n^C} (-1)^{|B_n|-|A_n^C|} \sum_{B_{-n}\geq A_{-n}^C} (-1)^{\sum_{t=1}^{n-1} |B_t|-|A_t^C|} p(x_{-n},B_{-n};x_n,B_n)=m(x,A)
\end{align*}
where the second equality follows from above.
\end{proof}

\subsection{Proof of Proposition \ref{p:SU_iff_assigns}}
\begin{proof}
(1) $\implies$ (2): Fix any $x \in A \in \mathcal{M}$. If $p(x,A)=\rho(x_1,A_1)\prod_{t=2}^n \rho_t(x_t,A_t|(A,x)^{t-1})=0$, then $\mu(C(x_1,A_1))=\rho_1(x_1,A_1)=0$ or $\mu(C(x_t,A_t)|C(A,x)^{t-1})=\rho_t(x_t,A_t|(A,x)^{t-1})=0$ for some $1<t\leq n$ with $(A,x)^{t-1} \in \mathcal{H}_{t-1}$, by definition of $p$.\footnote{If $\rho_1(x_1,A_1)>0$, then at least one term in the product $\prod_{t=2}^n \rho_t(x_t,A_t|(A,x)^{t-1})$ is zero. Let $t=\min\{1<s\leq n: \rho_s(x_s,A_s|(A,x)^{s-1})=0\}$: then for all $1<s<t$, $\rho_s(x_s,A_s|(A,x)^{s-1})>0$ and $(A,x)^{s-1} \in \mathcal{H}_{s-1}$, by definition of $\mathcal{H}_{s-1}$. Hence, $(A,x)^{t-1} \in \mathcal{H}_{t-1}$.}

If the former, then $C(x,A)\subseteq C(x_1,A_1)$ implies $\mu(C(x,A))\leq \mu(C(x_1,A_1))=0$. If the latter, then $C(x,A)\subseteq \bigcap_{s=1}^t C(x_s,A_s)$ implies $\mu(C(x,A))\leq \mu\big(\bigcap_{s=1}^t C(x_s,A_s)\big)=0$.\footnote{If $(A,x)^{t-1} \in \mathcal{H}_{t-1}$, then $\rho_s(x_s,A_s|(A,x)^{s-1})>0$ for all $1<s<t$ and $\rho_1(x_1,A_1)>0$, by definition of $\mathcal{H}_{t-1}$. I claim that $\mu(C(A,x)^s)>0$ for all $1<s\leq t-1$. Base case: $\mu(C(A,x)^2)=\mu(C(x_1,A_1)\cap C(x_2,A_2))=\mu(C(x_2,A_2)|C(x_1,A_1))\mu(C(x_1,A_1))=\rho_2(x_2,A_2|A_1,x_1)\rho_1(x_1,A_1)>0$. Inductive step: suppose $\mu(C(A,x)^s)>0$ for $1<s<t-1$. Since $(A,x)^{t-1} \in \mathcal{H}_{t-1}$, $(A,x)^s \in \mathcal{H}_s$. Hence, $\mu(C(A,x)^{s+1})=\mu(C(A,x)^s \cap C(x_{s+1},A_{s+1}))=\mu(C(x_{s+1},A_{s+1})|C(A,x)^s)\mu(C(A,x)^s)=\rho_{s+1}(x_{s+1},A_{s+1}|(A,x)^s)\mu(C(A,x)^s)>0$. Finally, since $\mu(C(A,x)^{t-1})>0$, we can write $\mu(\bigcap_{s=1}^t C(x_s,A_s))=\mu(C(x_t,A_t)|C(A,x)^{t-1})\mu(C(A,x)^{t-1})=\rho_t(x_t,A_t|(A,x)^{t-1})\mu(C(A,x)^{t-1})=0$.} In either case, $\mu(C(x,A))=0=p(x,A)$, as desired. If $\rho(x_1,A_1)\prod_{t=2}^n \rho_t(x_t,A_t|(A,x)^{t-1})=p(x,A)>0$, then every multiplicand of that product is positive. Hence,\footnote{By the previous footnote, $\mu(C(A,x)^{t-1})>0$ for all $1<t\leq n$.} 
$$\mu(C(x,A))=\mu(C(x_1,A_1))\prod_{t=2}^n \mu(C(x_t,A_t)|C(A,x)^{t-1})=\rho(x_1,A_1)\prod_{t=2}^n \rho_t(x_t,A_t|(A,x)^{t-1})=p(x,A)$$
(2) $\implies$ (3): Fix any $A<<X$ and $x \in A^C$. Since $x_t \succ_t A_t^C\backslash\{x_t\}$ if and only if $B_t^C \succ_t x_t \succ_t B_t\backslash\{x_t\}$ for some $B_t \supseteq A_t^C$,
$$C(x,A^C)=\bigcup_{B\geq A^C} E(x,B^C)$$
and this union is disjoint. Hence,
$$p(x,A^C)=\sum_{B\geq A^C} \mu(E(x,B^C))$$
By \textbf{Lemma \ref{l:inv}} with $f(B)=\mu(E(x,B^C))$ and $F(A)=p(x,A)$,
$$\mu(E(x,A))=f(A^C)=\sum_{B\geq A^C} m_L(A^C,B)p(x,B)=m(x,A)$$
(3) $\implies$ (1): For all $y \in D \in \mathcal{M}$, $y_t \succ_t D_t\backslash\{y_t\}$ if and only if $B_t \succ_t y_t \succ_t B_t^C\backslash\{y_t\}$ for some $B_t \subseteq D_t^C$ implies
$$C(y,D)=\bigcup_{B\leq D^C} E(y,B)$$
and this union is disjoint. By \textbf{Lemma \ref{p_BM_sums}},
$$\mu(C(y,D))=\sum_{B \leq D^C} m(y,B)=p(y,D)$$
Fix any $x_1 \in A_1 \in \mathcal{M}_1$ and fix $A_{-1} \in \mathcal{M}_{-1}$: since $\{C(x_1,A_1;x_{-1},A_{-1})\}_{x_{-1} \in A_{-1}}$ partitions $C(x_1,A_1)$,
$$\mu(C(x_1,A_1))=\sum_{x_{-1} \in A_{-1}} \mu(C(x_1,A_1;x_{-1},A_{-1}))=\sum_{x_{-1} \in A_{-1}} p(x_1,A_1;x_{-1},A_{-1})=\rho_1(x_1,A_1)$$
In particular, $\mu(C(h_1))=\rho_1(x_1,A_1)>0$ for any $h_1=(A_1,x_1) \in \mathcal{H}_1$, by definition of $\mathcal{H}_1$. Next, fix any $2<t\leq n$, $h_{t-1} \in \mathcal{H}_{t-1}$, and $A_{\geq t} \in \mathcal{M}_{\geq t}$. Since $\{C(h_{t-1};x_{\geq t},A_{\geq t})\}_{x_{\geq t} \in A_{\geq t}}$ partitions $C(h_{t-1})$,
\begin{align*}
    \mu(C(h_{t-1}))=\sum_{x_{\geq t} \in A_{\geq t}} \mu(C(h_{t-1};x_{\geq t},A_{\geq t}))=\sum_{x_{\geq t} \in A_{\geq t}} p(h_{t-1};x_{\geq t},A_{\geq t}) \\
    =p_{t-1}(h_{t-1})=\rho_1(x_1,A_1)\prod_{s=2}^{t-1} \rho_s(x_s,A_s|(A,x)^{s-1})>0
\end{align*}
by definition of $\mathcal{H}_{t-1}$. Hence, for all $1<t\leq n$, $h_{t-1}=(A,x)^{t-1} \in \mathcal{H}_{t-1}$, and $x_t \in A_t \in \mathcal{M}_t$,
\begin{align*}
    \mu(C(x_t,A_t)|C(h_{t-1}))=\frac{\mu(C(h_{t-1};x_t,A_t))}{\mu(C(h_{t-1}))} \\
    =\frac{\rho_1(x_1,A_1)\prod_{s=2}^t \rho_s(x_s,A_s|(A,x)^{s-1})}{\rho_1(x_1,A_1)\prod_{s=2}^{t-1} \rho_s(x_s,A_s|(A,x)^{s-1})}=\rho_t(x_t,A_t|h_{t-1})
\end{align*}
\end{proof}

\subsection{Proof of Theorem \ref{p:all3}}
\begin{proof}
$(\implies)$: As I have shown, $p$ satisfies Marginal Consistency by \textbf{Proposition \ref{p:SU_iff_assigns}}. To see that $p$ satisfies Joint Supermodularity, fix $y_t \neq x_t \in X_t$ for each $1 \leq t \leq n$. By \textbf{Proposition \ref{p:SU_iff_assigns}},
\begin{align*}
    \sum_{B\geq \{x\}^C: \sum_{t=1}^n |B_t| \text{even}} p(y,B)-\sum_{B\geq \{x\}^C: \sum_{t=1}^n |B_t| \text{odd}} p(y,B) \\
    =\sum_{B \geq \{y,z\}_N} (-1)^{\sum_{t=1}^n |B_t|-2n} p(x,B)=m(y,\{x\})=\mu(E(y,\{x\}))\geq 0
\end{align*}
$(\impliedby)$: Define $\mu(xyz):=m(y,\{x\}) \geq 0$. Observe that 
\begin{align*}
    \sum_{\succ \in P} \mu(\succ)=\sum_{d \in X} \bigg[\sum_{e \in (X\backslash\{d\})} m(e,\{d\})\bigg]=\sum_{d \in X} m(d,\emptyset)=\sum_{d \in X} p(d,X)=1
\end{align*}
where $(X\backslash\{d\})=(X_t\backslash\{d_t\})_{1\leq t\leq n}$, the first equality follows from counting (since there is a bijection between preference tuples and their first- and second-ranked elements in each component), the second equality follows from \textbf{Corollary \ref{c:sums_BM}}, and the third equality follows by definition of $m$. Hence, $\mu$ is a probability measure over $P$. Next, fix any $A<<X$ and $x \in A^C$. For $i=0,1,2$, let $T_i:=\{t \in \{1,\ldots,n\}: |A_t|=i\}$. Since $A<<X$, $\{T_0,T_1,T_2\}$ partition $\{1,\ldots,n\}$. Since 
$$E(x,A)=\bigcup_{y_{T_0}\neq x_{T_0}} \bigcup_{y_{T_2} \in A_{T_2}} E(y_{T_0},\{x\}_{T_0};x_{T_1},A_{T_1};y_{T_2},(A\backslash\{y\})_{T_2})$$
and this union is disjoint,
\begin{align*}
    \mu(E(x,A))=\sum_{y_{T_0}\neq x_{T_0}} \sum_{y_{T_2} \in A_{T_2}} m(y_{T_0},\{x\}_{T_0};x_{T_1},A_{T_1};y_{T_2},(A\backslash\{y\})_{T_2})=m(x,A)
\end{align*}
where the second equality follows from \textbf{Corollary \ref{c:sums_BM}}. By \textbf{Proposition \ref{p:SU_iff_assigns}}, $\mu$ is an SU representation of $\rho$. Finally, suppose $\mu'$ is an SU representation of $\rho$. Fix any $xyz \ \in P$: by \textbf{Proposition \ref{p:SU_iff_assigns}},
$$\mu(xyz)=\mu(E(y,\{x\}))=m(y,\{x\})=\mu(xyz)$$
Hence, $\mu$ is unique.
\end{proof}

\subsection{Proof of Theorem \ref{thm}}
\begin{proof}
$(\implies)$: By \textbf{Proposition \ref{p:SU_iff_assigns}}, Joint BM Nonnegativity and Marginal Consistency are necessary for SU, since $m(x,A)=\mu(E(x,A))\geq 0$ for all $A<<X$ and $x \in A^c$, and $\sum_{x_t \in A_t} p(x_t,A_t;x_{-t},A_{-t})=\mu(C(x_{-t},A_{-t}))$ is constant in $A_t$ for all $1\leq t<n$ and $x_{-t} \in A_{-t} \in \mathcal{M}_{-t}$.

$(\impliedby)$: First, I define the $t$-cylinders.
\begin{definition}
For $1\leq k\leq|X_t|$, a $\boldsymbol{k}$\textbf{-sequence} is a (distinct) sequence of elements $(x_t^1,\ldots,x_t^k)$ in $X_t$ and its $\boldsymbol{t}$\textbf{-cylinder}\footnote{This definition is analogous to \cite{chambers2016revealed}'s Chapter 7 definition of cylinders.} is
\begin{align*}
    I_{x_t^1,\ldots,x_t^k}:=\big\{\succ_t \ \in P_t: x_t^1 \succ_t \cdots \succ_t x_t^k \succ_t X_t\backslash \{x_t^1,\ldots,x_t^k\}\big\}
\end{align*}
\end{definition}

A $k$-sequence's $t$-cylinder is the set of period-$t$ preferences that rank that $k$-sequence in order above all other elements. Let $\mathcal{I}_t$ be the set of all $t$-cylinders, and let $\mathcal{I}:=\times_{t=1}^n \mathcal{I}_t$. Observe that $\mathcal{I}_t$ contains all singletons, since $I_{x_t^1,\ldots,x_t^{|X_t|-1}}=I_{x_t^1,\ldots,x_t^{|X_t|}}=\{\succ_t\}$ for the unique $\succ_t$ satisfying $x_t^1 \succ_t \cdots \succ_t x_t^{|X_t|}$. For $1 \leq t<n$, observe that $\mathcal{I}_t$ only contains $t$-cylinders of $k$-sequences for $k=1,2,3$. Given (nonempty) menu $A_t$, let $\pi(A_t)$ be the set of all $|A_t|$-sequences that permute $A_t$. Now, I recursively define $\nu: \mathcal{I} \rightarrow \mathbb{R}_{\geq 0}$.\footnote{My definition of $\nu$ is the multiperiod analog of \cite{chambers2016revealed} (7.4).} First, define
\begin{align*}
    \nu(I_{x_{-n}^1,x_{-n}^2} \times I_{x_n^1}):=m(x_{-n}^2,\{x\}_{-n}^1;x_n^1,\emptyset)
\end{align*}
Next, for $1<k\leq|X_n|$ and $x_n^K:=(x_n^1,\ldots,x_n^k)$, let $A_n=\{x_n^1,\ldots,x_n^{k-1}\}$ and define
\begin{align*}
    \nu(I_{x_{-n}^1,x_{-n}^2} \times I_{x_n^K}):=\begin{cases} 
      0 & \sum_{\tau_n \in \pi(A_n)} \nu(I_{x_{-n}^1,x_{-n}^2} \times I_{\tau_n})=0 \\
      \frac{\nu(I_{x_{-n}^1,x_{-n}^2} \times I_{x_n^{K-1}})m(x_{-n}^2,\{x\}_{-n}^1;x_n^k,A_j)}{\sum_{\tau_n \in \pi(A_n)} \nu(I_{x_{-n}^1,x_{-n}^2} \times I_{\tau_n})} & \text{else} 
   \end{cases}
\end{align*}
Finally, for each $I \in \mathcal{I}$, let $T_1=\{1 \leq t \leq n-1: I_t=I_{x_t^1}\}$ and $T_3=\{1 \leq t \leq n-1: I_t=I_{x_t^1,x_t^2,x_t^3}\}$, and define
\begin{align*}
    \nu(I):=\sum_{x_{T_1}^2\neq\neq x_{T_1}^1} \nu(I_{x_{T_1}^1,x_{T_1}^2} \times I_{x_{T_3}^1,x_{T_3}^2} \times I_{-(T_1\cup T_3)})
\end{align*}
Note that Joint BM Nonnegativity implies $\nu\geq 0$.

\begin{definition}
For any $k=(k_1,\ldots,k_n)$ with $0 \leq k_t<|X_t|$, the \textbf{first additive property $\boldsymbol{p_1(k)}$} holds if for all $A$ with $|A_t|=k_t$ and all $x_t \in A_t^C$,
\begin{align*}
    \sum_{\tau \in \times_{t \in -T} \pi(A_t)} \nu(I_{\tau,x_{-T}} \times I_{x_T})=m(x_{-T},A_{-T}; x_T,\emptyset)
\end{align*}
where $T=\{1 \leq t \leq n: k_t=0\}$.
\end{definition}

\begin{claim}\label{claim: p1k}
$p_1(k)$ holds for all $k$ with $0\leq k_t<|X_t|$ for each $1 \leq t \leq n$.
\end{claim}

\begin{proof}
Base case ($0\leq k_t<3$ for all $1 \leq t<n$, $k_n=0$): Fix any $A$ with $|A_t|=k_t$ for $1 \leq t<n$ and $A_n=\emptyset$, and fix any $x \in A^C$. For $0\leq i<3$, let $T_i=\{1\leq t<n: k_t=i\}$. Label $x_{T_0}^1:=x_{T_0}$, $A_{T_1}=\{x^1\}_{T_1}$, $x_{T_1}^2:=x_{T_1}$, $A_{T_2}=\{x^1,x^2\}_{T_2}$, and $x_{T_2}^3:=x_{T_2}$. Then

\begin{align*}
    \sum_{\tau_{T_2}\in\{x^1x^2,x^2x^1\}_{T_2}} \nu(I_{x_{T_0}^1} \times I_{x_{T_1}^1,x_{T_1}^2} \times I_{\tau_{T_2},x_{T_2}^3} \times I_{x_n}) \\
    =\sum_{\tau_{T_2}\in\{x^1x^2,x^2x^1\}_{T_2}} \sum_{x_{T_0}^2 \neq\neq x_{T_0}^1} \nu(I_{x_{T_0}^1,x_{T_0}^2} \times I_{x_{T_1}^1,x_{T_1}^2} \times I_{\tau_{T_2}} \times I_{x_n}) \\
    =\sum_{e_{T_0} \in \{x^1\}_{T_0}^C} \sum_{e_{T_2} \in A_{T_2}} m(e_{T_0},\{x^1\}_{T_0}; x_{T_1}^2, \{x^2\}_{T_1}; e_{T_2},(A\backslash\{e\})_{T_2}; x_n,\emptyset) \\
    =m(x_{T_0}^1,\emptyset;x_{T_1}^2, \{x^1\}_{T_1};x_{T_2}^3,A_{T_2};x_n,\emptyset)
\end{align*}
where the last equality follows from \textbf{Corollary \ref{c:sums_BM}}.

First inductive step ($k_t=1$ for all $1\leq t<n$, $0<k_n<|X_n|$): Fix any $A$ with $A_t=\{x_t^1\}$ for all $1\leq t<n$ and $A_n=\{x_n^1,\ldots,x_n^{k_n}\}$. Fix any $x \in A^C$. First, observe that
\begin{align*}
    \sum_{\tau_n \in \pi(A_n)} \nu(I_{x_{-n}^1,x_{-n}} \times I_{\tau_n})=\sum_{y_n \in A_n} \bigg[\sum_{\tau \in A_n\backslash\{y_n\}} \nu(I_{x_{-n}^1,x_{-n}} \times I_{\tau,y_n})\bigg] \\
    =\sum_{y_n \in A_n} m(x_{-n},\{x^1\}_{-n};y_n,A_n\backslash\{y_n\})=\sum_{x_n \in A_n^C} m(x_{-n},\{x^1\}_{-n};x_n,A_n)
\end{align*}
where the first equality holds because permuting $A_n$ is equivalent to choosing the last element and permuting the remaining $k_n-1$ elements, the second equality holds by the inductive hypothesis $p(k_{-n},k_n-1)$,\footnote{Strictly speaking, for $k_n=1$ the inner sum on the first line is not well-defined. However, in this case it still holds that $\nu(I_{x_{-n}^1,x_{-n}} \times I_{x_n^1})=m(x_{-n},\{x^1\}_{-n};x_n^1,\emptyset)=\sum_{x_n \neq x_n^1} m(x_{-n},\{x^1\}_{-n};x_n,\{x_n^1\})$, where the first equality follows by $p(k_{-n},0)$ (or directly by definition of $\nu$).} and the third equality holds by \textbf{Lemma \ref{l:sums_BM_sums}}. There are two cases.
\begin{enumerate}
    \item If $\sum_{\tau_n \in \pi(A_n)} \nu(I_{x_{-n}^1,x_{-n}} \times I_{\tau_n})=0$, then for each $\tau_n \in \pi(A_n)$ it follows by definition of $\nu$ that $\nu(I_{x_{-n}^1,x_{-n}} \times I_{\tau_n,x_n})=0$. Furthermore, Joint BM Nonnegativity implies $m(x_{-n},\{x^1\}_{-n};x_n,A_n)=0$ for each $x_n \in A_n^C$. Hence,
    \begin{align*}
        \sum_{\tau_n \in \pi(A_n)} \nu(I_{x_{-n}^1,x_{-n}} \times I_{\tau_n,x_n})=0=m(x_{-n},\{x^1\}_{-n};x_n,A_n)
    \end{align*}
    as desired.
    \item If $\sum_{\tau_n \in \pi(A_n)} \nu(I_{x_{-n}^1,x_{-n}} \times I_{\tau_n})>0$, then for each $\tau_n \in \pi(A_n)$ it follows by definition of $\nu$ that
    \begin{align*}
        \sum_{\tau_n \in \pi(A_n)} \nu(I_{x_{-n}^1,x_{-n}} \times I_{\tau_n,x_n})=\sum_{\tau_n \in \pi(A_n)}\Bigg[\frac{\nu(I_{x_{-n}^1,x_{-n}} \times I_{\tau_n})m(x_{-n},\{x^1\}_{-n};x_n,A_n)}{\sum_{\tau_n' \in \pi(A_n)} \nu(I_{x_{-n}^1,x_{-n}} \times I_{\tau_n'})}\Bigg] \\
        =m(x_{-n},\{x^1\}_{-n};x_n,A_n)
    \end{align*}
    as desired.
\end{enumerate}

Second inductive step ($0\leq k_t<3$ for all $1 \leq t<n$, $k_t\neq 1$ for some $1 \leq t<n$, $0<k_n<|X_n|$): Fix any $A$ with $|A_t|=k_t$ for $1 \leq t\leq n$ and any $x \in A^C$. For $0\leq i<3$, let $T_i=\{1\leq t<n: k_t=i\}$. By assumption, $A_{T_0}=\emptyset$. Label $x_{T_0}^1:=x_{T_0}$, $A_{T_1}=\{x^1\}_{T_1}$, $x_{T_1}^2:=x_{T_1}$, $A_{T_2}=\{x^1,x^2\}_{T_2}$, $x_{T_2}^3:=x_{T_2}$, and $A_n=\{x_n^1,\ldots,x_n^{k_n}\}$. First, observe that
\begin{align*}
    \sum_{\tau_{T_2} \in \{x^1x^2,x^2x^1\}_{T_2}}
    \sum_{\tau_n \in \pi(A_n)}\nu(I_{x_{T_0}^1} \times I_{x_{T_1}^1,x_{T_1}^2} \times I_{\tau_{T_2},x_{T_2}^3} \times I_{\tau_n}) \\
    =\sum_{y_n \in A_n} \bigg[\sum_{\tau_{T_2} \in \{x^1x^2,x^2x^1\}_{T_2}} \sum_{\tau \in \pi(A_n\backslash\{y_n\})} \nu(I_{x_{T_0}^1} \times I_{x_{T_1}^1,x_{T_1}^2} \times I_{\tau_{T_2},x_{T_2}^3} \times I_{\tau,y_n})\bigg] \\
    =\sum_{y_n \in A_n} m(x_{T_0}^1,\emptyset;x_{T_1}^2,\{x^1\}_{T_1};x_{T_2}^3,A_{T_2};y_n,A_n\backslash\{y_n\})=\sum_{x_n \in A_n^C} m(x_{T_0}^1,\emptyset;x_{T_1}^2,\{x^1\}_{T_1};x_{T_2}^3,A_{T_2};x_n,A_n)
\end{align*}
where the first equality holds because permuting $A_n$ is equivalent to choosing the last element and permuting the remaining $k_n-1$ elements, the second equality holds by the inductive hypothesis $p_1(k_{-n},k_n-1)$,\footnote{Strictly speaking, for $k_n=1$ the inner sum on the second line is not well-defined. However, in this case it still holds that $\sum_{\tau_{T_2} \in \{x^1x^2,x^2x^1\}_{T_2}} \nu(I_{x_{T_0}^1} \times I_{x_{T_1}^1,x_{T_1}^2} \times I_{\tau_{T_2},x_{T_2}^3} \times I_{x_n^1})=m(x_{T_0}^1,\emptyset;x_{T_1}^2,\{x^1\}_{T_1};x_{T_2}^3,A_{T_2};x_n^1,\emptyset)=\sum_{x_n \neq x_n^1} m(x_{T_0}^1,\emptyset;x_{T_1}^2,\{x^1\}_{T_1};x_{T_2}^3,A_{T_2};x_n,\{x_n^1\})$, where the first equality follows from $p(k_{-n},0$, as shown in the base case.} and the third equality holds because of \textbf{Lemma \ref{l:sums_BM_sums}}. There are two cases. 

\begin{enumerate}
    \item If
    \begin{align*}
        \sum_{\tau_{T_2} \in \{x^1x^2,x^2x^1\}_{T_2}} \sum_{\tau_n \in \pi(A_n)}\nu(I_{x_{T_0}^1} \times I_{x_{T_1}^1,x_{T_1}^2} \times I_{\tau_{T_2},x_{T_2}^3} \times I_{\tau_n}) \\
        =\sum_{\tau_{T_2} \in \{x^1x^2,x^2x^1\}_{T_2}} \sum_{x_{T_0}^2\neq\neq x_{T_0}^1} \sum_{\tau_n \in \pi(A_n)}\nu(I_{x_{T_0}^1,x_{T_0}^2} \times I_{x_{T_1}^1,x_{T_1}^2} \times I_{\tau_{T_2}} \times I_{\tau_n})=0
    \end{align*}
    where the second equality follows by definition of $\nu$, then $\nu \geq 0$ implies that for each $x_{T_0}^2\neq\neq x_{T_0}^1$ and $\tau_{T_2} \in \{x^1x^2,x^2x^1\}_{T_2}$, $\sum_{\tau_n \in \pi(A_n)}\nu(I_{x_{T_0}^1,x_{T_0}^2} \times I_{x_{T_1}^1,x_{T_1}^2} \times I_{\tau_{T_2}} \times I_{\tau_n})=0$. By definition of $\nu$, it follows that $\nu(I_{x_{T_0}^1,x_{T_0}^2} \times I_{x_{T_1}^1,x_{T_1}^2} \times I_{\tau_{T_2}} \times I_{\tau_n,x_n})=0$. Furthermore, Joint BM Nonnegativity implies $m(x_{T_0}^1,\emptyset;x_{T_1}^2,\{x^1\}_{T_1};x_{T_2}^3,A_{T_2};x_n,A_n)=0$ for each $x_n \in A_n^C$. Hence,
    \begin{align*}
    \sum_{\tau_{T_2} \in \{x^1x^2,x^2x^1\}_{T_2}}
    \sum_{\tau_n \in \pi(A_n)}\nu(I_{x_{T_0}^1} \times I_{x_{T_1}^1,x_{T_1}^2} \times I_{\tau_{T_2},x_{T_2}^3} \times I_{\tau_n,x_n}) \\
    =\sum_{\tau_{T_2} \in \{x^1x^2,x^2x^1\}_{T_2}} \sum_{x_{T_0}^2\neq\neq x_{T_0}^1}
    \sum_{\tau_n \in \pi(A_n)}\nu(I_{x_{T_0}^1,x_{T_0}^2} \times I_{x_{T_1}^1,x_{T_1}^2} \times I_{\tau_{T_2}} \times I_{\tau_n,x_n}) \\
    =0=m(x_{T_0}^1,\emptyset;x_{T_1}^2,\{x^1\}_{T_1};x_{T_2}^3,A_{T_2};x_n,A_n)
    \end{align*}
    as desired.
    \item If
    $$\sum_{\tau_{T_2} \in \{x^1x^2,x^2x^1\}_{T_2}} \sum_{x_{T_0}^2\neq\neq x_{T_0}^1} \sum_{\tau_n \in \pi(A_n)}\nu(I_{x_{T_0}^1,x_{T_0}^2} \times I_{x_{T_1}^1,x_{T_1}^2} \times I_{\tau_{T_2}} \times I_{\tau_n})>0$$
    Let $T_>=\{\tau_{T_2} \in \{x^1x^2,x^2x^1\}_{T_2},x_{T_0}^2\neq\neq x_{T_0}^1: \sum_{\tau_n \in \pi(A_n)}\nu(I_{x_{T_0}^1,x_{T_0}^2} \times I_{x_{T_1}^1,x_{T_1}^2} \times I_{\tau_{T_2}} \times I_{\tau_n})>0\}$ and $T=\{\tau_{T_2} \in \{x^1x^2,x^2x^1\}_{T_2},x_{T_0}^2\neq\neq x_{T_0}^1: \sum_{\tau_n \in \pi(A_n)}\nu(I_{x_{T_0}^1,x_{T_0}^2} \times I_{x_{T_1}^1,x_{T_1}^2} \times I_{\tau_{T_2}} \times I_{\tau_n})=0\}$. For $(\tau_{T_2},x_{T_0}^2) \in T$, it follows that $\nu(I_{x_{T_0}^1,x_{T_0}^2} \times I_{x_{T_1}^1,x_{T_1}^2} \times I_{\tau_{T_2}} \times I_{\tau_n,x_n})=0$ for each $\tau_n \in \pi(A_n)$ and therefore
    \begin{align*}
        m((x^2,\{x^1\})_{T_0\cup T_1};\tau_{T_2,2},\{\tau_1\}_{T_2};x_n,A_n)=\sum_{\tau_n \in \pi(A_n)} \nu(I_{x_{T_0}^1,x_{T_0}^2} \times I_{x_{T_1}^1,x_{T_1}^2} \times I_{\tau_{T_2}} \times I_{\tau_n,x_n})=0
    \end{align*}
    where the first equality follows by $p(1,\ldots,1,k_n)$. Hence,
\begin{align*}
    \sum_{\tau_{T_2} \in \{x^1x^2,x^2x^1\}_{T_2}}
    \sum_{\tau_n \in \pi(A_n)}\nu(I_{x_{T_0}^1} \times I_{x_{T_1}^1,x_{T_1}^2} \times I_{\tau_{T_2},x_{T_2}^3} \times I_{\tau_n,x_n}) \\
    =\sum_{\tau_{T_2} \in \{x^1x^2,x^2x^1\}_{T_2}} \sum_{x_{T_0}^2\neq\neq x_{T_0}^1}
    \sum_{\tau_n \in \pi(A_n)}\nu(I_{x_{T_0}^1,x_{T_0}^2} \times I_{x_{T_1}^1,x_{T_1}^2} \times I_{\tau_{T_2}} \times I_{\tau_n,x_n}) \\
    =\sum_{(\tau_{T_2},x_{T_0}^2) \in T_>} \sum_{\tau_n \in \pi(A_n)}\nu(I_{x_{T_0}^1,x_{T_0}^2} \times I_{x_{T_1}^1,x_{T_1}^2} \times I_{\tau_{T_2}} \times I_{\tau_n,x_n}) \\
    =\sum_{(\tau_{T_2},x_{T_0}^2) \in T_>} \sum_{\tau_n \in \pi(A_n)} \frac{\nu(I_{x_{T_0}^1,x_{T_0}^2} \times I_{x_{T_1}^1,x_{T_1}^2} \times I_{\tau_{T_2}} \times I_{\tau_n}) m((x^2,\{x^1\})_{T_0\cup T_1};\tau_{T_2,2},\{\tau_1\}_{T_2};x_n,A_n)}{\sum_{\tau_n' \in \pi(A_n)} \nu(I_{x_{T_0}^1,x_{T_0}^2} \times I_{x_{T_1}^1,x_{T_1}^2} \times I_{\tau_{T_2}} \times I_{\tau_n'})} \\
    =\sum_{(\tau_{T_2},x_{T_0}^2) \in T_>} m((x^2,\{x^1\})_{T_0\cup T_1};\tau_{T_2,2},\{\tau_1\}_{T_2};x_n,A_n) \\
    =\sum_{\tau_{T_2} \in \{x^1x^2,x^2x^1\}_{T_2}} \sum_{x_{T_0}^2\neq\neq x_{T_0}^1} m((x^2,\{x^1\})_{T_0\cup T_1};\tau_{T_2,2},\{\tau_1\}_{T_2};x_n,A_n) \\
    =m(x_{T_0}^1,\emptyset;x_{T_1}^2,\{x^1\}_{T_1};x_{T_2}^3,A_{T_2};x_n,A_n)
\end{align*}
as desired.
\end{enumerate}
\end{proof}

\begin{definition}
For any $1 \leq t \leq n$ and $k=(k_1,\ldots,k_n)$ with $0<k_s \leq |X_s|$ and $k_t<|X_t|$, the \textbf{second additive property $\boldsymbol{p_2(t,k)}$} holds if for all $A$ with $|A_s|=k_s$ and $\tau \in \times_{s=1}^n \pi(A_s)$,
$$\sum_{x_t \in A_t^C} \nu(I_{\tau_{-t}} \times I_{\tau_t,x_t})=\nu(I_{\tau_{-t}} \times I_{\tau_t})$$
\end{definition}

\begin{claim}
$p_2(t,k)$ holds for all $1 \leq t \leq n$ and $k$ with $0<k_s \leq |X_s|$ and $k_t<|X_t|$.
\end{claim}

\begin{proof}
Fix any $1 \leq t \leq n$ and $k$ with $0<k_s \leq |X_s|$ and $k_t<|X_t|$. The following cases are exhaustive.
\begin{enumerate}
    \item ($t=n$ and $k_s=2$ for all $1\leq s<n$): Fix any $A$ with $|A_s|=2$ for $1 \leq s<n$ and $|A_n|=k_n$, and fix $\tau \in \times_{s=1}^n \pi(A_s)$. Label $\tau_s=x_s^1,x_s^2$ for $1 \leq s<n$. There are two subcases: if
    $$\sum_{\tau_n' \in \pi(A_n)} \nu(I_{x_{-n}^1,x_{-n}^2} \times I_{\tau_n'})=0$$
    then $\nu(I_{x_{-n}^1,x_{-n}^2} \times I_{\tau_n,x_n})=0$ for each $x_n \in A_n^C$ by definition of $\nu$ and $\nu(I_{x_{-n}^1,x_{-n}^2} \times I_{\tau_n'})=0$ for each $\tau_n' \in \pi(A_n)$ since $\nu\geq 0$. Hence,
    \begin{align*}
        \sum_{x_n \in A_n^C} \nu(I_{x_
        {-n}^1,x_{-n}^2} \times I_{\tau_n,x_n})=0=\nu(I_{x_{-n}^1,x_{-n}^2} \times I_{\tau_n})
    \end{align*}
    as desired. Else, 
    \begin{align*}
        \sum_{x_n \in A_n^C} \nu(I_{x_
        {-n}^1,x_{-n}^2} \times I_{\tau_n,x_n})=\sum_{x_n \in A_n^C} \Bigg[\frac{\nu(I_{x_{-n}^1,x_{-n}^2} \times I_{\tau_n})m(x_{-n}^2,\{x\}_{-n}^1;x_n,A_n)}{\sum_{\tau_n' \in \pi(A_n)} \nu(I_{x_{-n}^1,x_{-n}^2} \times I_{\tau_n'})}\Bigg] \\
        =\nu(I_{x_{-n}^1,x_{-n}^2} \times I_{\tau_n})\Bigg[\frac{\sum_{x_n \in A_n^C} m(x_{-n}^2,\{x\}_{-n}^1;x_n,A_n)}{\sum_{\tau_n' \in \pi(A_n)} \nu(I_{x_{-n}^1,x_{-n}^2} \times I_{\tau_n'})}\Bigg] \\
        =\nu(I_{x_{-n}^1,x_{-n}^2} \times I_{\tau_n})
    \end{align*}
    where the last equality follows from $p_1(1,\ldots,1,k_n-1)$.\footnote{Note that $0\leq k_n-1<|X_n|$. See the proof of the first inductive step of \textbf{Claim \ref{claim: p1k}}.}
    \item ($t=n$ and $k_s\neq 2$ for some $1\leq s<n$): Fix any $A$ with $|A_s|=k_s$ for all $1 \leq s \leq n$, and fix $\tau \in \times_{s=1}^n \pi(A_s)$. For $i=1,2,3$, let $S_i=\{1 \leq s<n: k_s=i\}$ and note that these form a partition of $\{1,\ldots,n-1\}$. Label $\tau_{S_1}=x_{S_1}^1$, $\tau_{S_2}=x_{S_2}^1,x_{S_2}^2$, and $\tau_{S_3}=x_{S_3}^1,x_{S_3}^2,x_{S_3}^3$. It follows that
    \begin{align*}
        \sum_{x_n \in A_n^C} \nu(I_{\tau_{-n}} \times I_{\tau_n,x_n})=\sum_{x_n \in A_n^C} \nu(I_{x_{S_1}^1} \times I_{x_{S_2}^1,x_{S_2}^2} \times I_{x_{S_3}^1,x_{S_3}^2,x_{S_3}^3} \times I_{\tau_n,x_n}) \\
        =\sum_{x_{S_1}^2 \neq\neq x_{S_1}^1} \sum_{x_n \in A_n^C} \nu(I_{x_{S_1}^1,x_{S_1}^2} \times I_{x_{S_2}^1,x_{S_2}^2} \times I_{x_{S_3}^1,x_{S_3}^2} \times I_{\tau_n,x_n}) \\
        =\sum_{x_{S_1}^2 \neq\neq x_{S_1}^1} \nu(I_{x_{S_1}^1,x_{S_1}^2} \times I_{x_{S_2}^1,x_{S_2}^2} \times I_{x_{S_3}^1,x_{S_3}^2} \times I_{\tau_n}) \\
        =\nu(I_{x_{S_1}^1} \times I_{x_{S_2}^1,x_{S_2}^2} \times I_{x_{S_3}^1,x_{S_3}^2,x_{S_3}^3} \times I_{\tau_n})=\nu(I_{\tau_{-n}} \times I_{\tau_n})
    \end{align*}
    where the first and last equalities follow by definition of $S_i$ and $\tau_{-n}$, the second and fourth follow by definition of $\nu$, and the third follows from my proof of $p_2(2,\ldots,2,k_n)$ from above.
    \item ($1\leq t<n$): Fix any $A$ with $|A_s|=k_s$ and $\tau \in \times_{s=1}^n \pi(A_s)$. For $i=1,2,3$, let $S_i=\{1 \leq s<n: k_s=i\}$ and note that these form a partition of $\{1,\ldots,n-1\}$. Label $\tau_{S_1}=x_{S_1}^1$, $\tau_{S_2}=x_{S_2}^1,x_{S_2}^2$, and $\tau_{S_3}=x_{S_3}^1,x_{S_3}^2,x_{S_3}^3$. Since $0<k_t<3$, there are two subcases. If $k_t=1$, label $\tau_t=x_t^1$. Then by definition of $\nu$,
    \begin{align*}
        \sum_{x_t \neq x_t^1} \nu(I_{\tau_{-t}} \times I_{x_t^1,x_t})=\sum_{x_t \neq x_t^1} \nu(I_{x_{S_1\backslash\{t\}}^1} \times I_{x_{S_2}^1,x_{S_2}^2} \times I_{x_{S_3}^1,x_{S_3}^2,x_{S_3}^3} \times I_{\tau_n} \times I_{x_t^1,x_t}) \\
        =\sum_{x_t \neq x_t^1} \sum_{x_{S_1\backslash\{t\}}^2 \neq\neq x_{S_1\backslash\{t\}}^1} \nu(I_{x_{S_1\backslash\{t\}}^1,x_{S_1S_1\backslash\{t\}}^2} \times I_{x_{S_2}^1,x_{S_2}^2} \times I_{x_{S_3}^1,x_{S_3}^2} \times I_{\tau_n} \times I_{x_t^1,x_t}) \\
        =\nu(I_{\tau_{-t}}\times I_{x_t^1})
    \end{align*}
    Similarly, if $k_t=2$ label $\tau_t=x_t^1,x_t^2$. Then by definition of $\nu$,
    \begin{align*}
        \nu(I_{\tau_{-t}} \times I_{x_t^1,x_t^2,x_t^3})=\nu(I_{x_{S_1}^1} \times I_{x_{S_2\backslash\{t\}}^1,x_{S_2\backslash\{t\}}^2} \times I_{x_{S_3}^1,x_{S_3}^2,x_{S_3}^3} \times I_{\tau_n} \times I_{x_t^1,x_t^2,x_t^3}) \\
        =\sum_{x_{S_1}^2 \neq\neq x_{S_1}^1} \nu(I_{x_{S_1}^1,x_{S_1}^2} \times I_{x_{S_2\backslash\{t\}}^1,x_{S_2\backslash\{t\}}^2} \times I_{x_{S_3}^1,x_{S_3}^2} \times I_{\tau_n} \times I_{x_t^1,x_t^2})=\nu(I_{\tau_{-t}} \times I_{x_t^1,x_t^2})
    \end{align*}
    as desired.
\end{enumerate}
\end{proof}
Next, define $\mu: 2^P\rightarrow \R$ as
\begin{align*}
    \mu(\succ):=\nu(I_{x_{-n}^1,x_{-n}^2,x_{-n}^3} \times I_{x_n^1,\ldots,x_n^{|X_n|}})
\end{align*}
for the (unique) $x_{-n}^1,x_{-n}^2,x_{-n}^3$ and $x_n^1,\ldots,x_n^{|X_n|}$ satisfying $x_t^1\succ_t x_t^2\succ_t x_t^3$ for all $1\leq t<n$ and $x_n^1\succ_n\cdots \succ_nx_n^{|X_n|}$, and $\mu(C)=\sum_{\succ \in C} \mu(\succ)$ for all $C \in 2^P$.

\begin{claim}
$\mu$ is a probability measure.
\end{claim}

\begin{proof}
Since $\mu\geq 0$, it suffices to show that $\sum_{\succ \in P} \mu(\succ)=1$. Rewriting this sum yields
\begin{align*}
    \sum_{\succ \in P} \mu(\succ)=\sum_{\tau \in \times \pi(X_t)} \nu(I_{\tau})=\sum_{y \in X} \sum_{\tau' \in \times \pi(X_t\backslash\{y_t\})} \nu(I_{\tau',y})=\sum_{y \in X} m(y,(X_t\backslash\{y_t\})) \\
    =\sum_{y \in X} \sum_{B\geq (\{y_t\})} (-1)^{\sum |B_t|-n} p(y,B)=\sum_{B:|B_t|\geq 1} (-1)^{\sum |B_t|-n} \sum_{y \in B} p(y,B)=\sum_{B:|B_t|\geq 1} (-1)^{\sum |B_t|-n} \\
    =\sum_{B:|B_t|\geq 1} \prod_{t=1}^n (-1)^{|B_t|-1}=\prod_{t=1}^n \sum_{B_t: |B_t|\geq 1} (-1)^{|B_t|-1}=\prod_{t=1}^n \sum_{k=1}^{|X_t|} (-1)^{k-1}\binom{|X_t|}{k}=1
\end{align*}
\end{proof}
where the second equality holds because permuting $X_t$ is equivalent to choosing the last element and permuting the remaining $|X_t|-1$ elements, the third equality holds by $p_1((|X_t|-1))$, the fifth equality follows from $y \in X$ and $B\geq (\{y_t\})$ iff $|B_t| \geq 1$ and $y \in B$, and the ninth equality follows from $1-\sum_{k=1}^n (-1)^{k-1} \binom{n}{k}=\sum_{k=0}^n (-1)^k \binom{n}{k}=(1-1)^n=0$ for all $n\geq 1$.

\begin{claim}\label{claim: extends}
$\mu$ extends $\nu$.
\end{claim}

\begin{proof}
The proof proceeds by induction on the length of $t$-cylinders' $k$-sequences. Base case ($k_t=2$ for all $1 \leq t<n$, $k_n=|X_n|$): for any $x_{-n}^1\neq\neq x_{-n}^2 \in X_{-n}$ and $\tau_n \in \pi(X_n)$,
\begin{align*}
    \mu(I_{x_{-n}^1,x_{-n}^2}\times I_{\tau_n})=\nu(I_{x_{-n}^1,x_{-n}^2,x_{-n}^3}\times I_{\tau_n})=\nu(I_{x_{-n}^1,x_{-n}^2}\times I_{\tau_n})
\end{align*}
where the first equality follows because $I_{x_{-n}^1,x_{-n}^2}\times I_{\tau_n}=\{\succ\}$ for the (unique) $\succ$ satisfying $x_{-n}^1\succ_{-n} x_{-n}^2\succ_{-n} x_{-n}^3$ and $\tau_n^1\succ_n\cdots\succ_n\tau_n^{|X_n|}$ and the second equality holds by definition of $\nu$.

First inductive step ($k_t=2$ for all $1 \leq t<n$, $1\leq k_n<|X_n|$): for any $x_{-n}^1\neq\neq x_{-n}^2 \in X_{-n}$, $A_n$ with $|A_n|=k_n$, and $\tau_n \in \pi(A_n)$,
\begin{align*}
    \mu(I_{x_{-n}^1,x_{-n}^2}\times I_{\tau_n})=\sum_{x_n \in A_n^C}\mu(I_{x_{-n}^1,x_{-n}^2}\times I_{\tau_n,x_n})=\sum_{x_n \in A_n^C} \nu(I_{x_{-n}^1,x_{-n}^2}\times I_{\tau_n,x_n})=\nu(I_{x_{-n}^1,x_{-n}^2}\times I_{\tau_n})
\end{align*}
where the first equality holds because $I_{x_{-n}^1,x_{-n}^2}\times I_{\tau_n}=\bigcup_{x_n \in A_n^C}\big(I_{x_{-n}^1,x_{-n}^2}\times I_{\tau_n,x_n}\big)$ holds and is a disjoint union, the second equality follows from the inductive hypothesis, and the third equality follows from $p_2(n,2,\ldots,2,k_n)$.

Second inductive step ($1\leq k_t\leq 3$ for all $1 \leq t<n$ with $k_t\neq 2$ for some $1\leq t<n$, $1 \leq k_n\leq |X_n|$): fix any $A$ with $|A_t|=k_t$ and any $\tau \in \times \pi(A_t)$. For $i=1,2,3$, let $T_i=\{1\leq t<n: k_t=i\}$ and note that these form a partition of $\{1,\ldots,n-1\}$. Label $\tau_{T_1}=x_{T_1}^1$, $\tau_{T_2}=x_{T_2}^1,x_{T_2}^2$, and $\tau_{T_3}=x_{T_3}^1,x_{T_3}^2,x_{T_3}^3$. Hence,
\begin{align*}
    \mu(I_\tau)=\mu(I_{x_{T_1}^1}\times I_{x_{T_2}^1,x_{T_2}^2}\times I_{x_{T_3}^1,x_{T_3}^2,x_{T_3}^3}\times I_{\tau_n})=\sum_{x_{T_1}^2\neq\neq x_{T_1}^1} \mu(I_{x_{T_1}^1,x_{T_1}^2}\times I_{x_{T_2}^1,x_{T_2}^2}\times I_{x_{T_3}^1,x_{T_3}^2}\times I_{\tau_n}) \\
    =\sum_{x_{T_1}^2\neq\neq x_{T_1}^1} \nu(I_{x_{T_1}^1,x_{T_1}^2}\times I_{x_{T_2}^1,x_{T_2}^2}\times I_{x_{T_3}^1,x_{T_3}^2}\times I_{\tau_n})=\nu(I_{\tau})
\end{align*}
where the second equality holds because $I_{x_{T_1}^1}\times I_{x_{T_2}^1,x_{T_2}^2}\times I_{x_{T_3}^1,x_{T_3}^2,x_{T_3}^3}\times I_{\tau_n}=\bigcup_{x_{T_1}^2\neq\neq x_{T_1}^1} I_{x_{T_1}^1,x_{T_1}^2}\times I_{x_{T_2}^1,x_{T_2}^2}\times I_{x_{T_3}^1,x_{T_3}^2}\times I_{\tau_n}$ holds and is a disjoint union, the third equality follows from the first inductive step, and the fourth equality follows by definition of $\nu$.
\end{proof}

\begin{claim}
$\mu(E(x,A))=m(x,A)$ for all $A<<X$ and $x \in A^C$.
\end{claim}

\begin{proof}
Fix any $A<<X$ and $x \in A^C$, and let $T=\{1\leq t\leq n: k_t=0\}$. Then
\begin{align*}
    \mu(E(x,A))=\sum_{\tau\in\times_{-T}\pi(A_t)} \mu(I_{\tau,x_{-T}} \times I_{x_T})=\sum_{\tau\in\times_{-T}\pi(A_t)} \nu(I_{\tau,x_{-T}} \times I_{x_T})=m(x,A)
\end{align*}
where the first equality holds because $E(x,A)=\bigcup_{\tau\in\times_{-T}\pi(A_t)}I_{\tau,x_{-T}} \times I_{x_T}$ holds and is a disjoint union, the second equality follows from \textbf{Claim \ref{claim: extends}}, and the third equality follows from $p_1(|A_1|,\ldots,|A_n|)$.
\end{proof}

By \textbf{Proposition \ref{p:SU_iff_assigns}}, I conclude $\mu$ is an SU representation of $\rho$.
\end{proof}

\subsection{Proof of Corollary \ref{c:fullsup}}
Define $\nu$ and $\mu$ as in the proof of \textbf{Theorem \ref{thm}}. Since Joint BM Positivity implies Joint BM Nonnegativity, $\mu$ is an SU representation. By construction, $\nu>0$, and hence $\mu>0$.

\subsection{Proof of Theorem \ref{SU4axioms}}
\begin{proof}
($\implies$): I have shown that SU implies Joint BM Nonnegativity and Marginal Consistency (\textbf{Theorem \ref{thm}}), Joint BM Nonnegativity implies Partial Marginal and Conditional BM Nonnegativity (directly following \textbf{Axioms \ref{ax:pmbmn} and \ref{ax:pcbmn}}), and Marginal Consistency implies $(-n)$-Marginal Consistency (directly following \textbf{Axiom \ref{ax:-nmc}}). Hence, it remains to show that SU implies $(-n)$-Conditional Consistency. 

Let $\mu$ be an SU representation and fix any $x_n \in A_n \in \mathcal{M}_n$ and $(A,x)^{-n} \in \mathcal{H}_{n-1}$ with partition $(y_{-n}^i,\{x^i\}_{-n})_{i \in I}$. Since $|X_t|=3$ for all $1 \leq t<n$, I can identify each index $i$ with the (unique) preference tuple $\succ_{-n}^i \in P_{-n}$ satisfying $E_{-n}(y_{-n}^i,\{x^i\}_{-n})=\{\succ_{-n}^i\}$. Let $(\succ_{-n}^i):=\{\succ_{-n}^i\}\times P_n=E(y^i,\{x^i\})^{-n}$ and define $I'$ as before: by \textbf{Proposition \ref{p:SU_iff_assigns}}, $I'=\{i\in I: \mu(\succ_{-n}^i)>0\}$, since $\mu(\succ_{-n}^i)=\sum_{x_n \in X_n} \mu((\succ_{-n}^i)\cap E(x_n,\emptyset))=\sum_{x_n \in X_n} m(y_{-n}^i,\{x^i\}_{-n};x_n,\emptyset)=m(y_{-n}^i,\{x^i\}_{-n})$. Hence,
\begin{align*}
    \rho_n(x_n,A_n|(A,x)^{-n})=\mu(C(x_n,A_n)|C(x,A)^{-n}) \\
    =\sum_{\succ_{-n} \in C(x,A)^{-n}: \mu(\succ_{-n})>0} \mu(C(x_n,A_n)|\succ_{-n})\mu(\succ_{-n}|C(x,A)^{-n}) \\
    =\sum_{i \in I'} \mu(C(x_n,A_n)|\succ_{-n}^i)\mu(\succ_{-n}^i|C(x,A)^{-n}) \\
    =\sum_{i \in I'} \frac{\sum_{D_n\subseteq A_n^C} \mu((\succ_{-n}^i) \cap E(x_n,D_n))}{\mu(\succ_{-n}^i)}\frac{\mu(\succ_{-n}^i)}{\mu(C(x,A)^{-n})} \\
    =\sum_{i \in I'} \rho_n(x_n,A_n|(y^i,\{x^i\})^{-n})\frac{m(y_{-n}^i,\{x^i\}_{-n})}{p_{-n}(x_{-n},A_{-n})}
\end{align*}
where the second equality follows from the Law of Total Probability, the fourth equality follows because $\bigcup_{D_n\subseteq A_n^C} ((\succ_{-n}^i)\cap E(x_n,D_n))=(\succ_{-n}^i)\cap \bigcup_{D_n\subseteq A_n^C}E(x_n,D_n)=(\succ_{-n}^i)\cap C(x_n,A_n)$ is a disjoint union, and the fifth equality follows by definition of $\rho_n(x_n,A_n|(y^i,\{x^i\})^{-n})$ and by \textbf{Proposition \ref{p:SU_iff_assigns}}, since $\mu((\succ_{-n}^i) \cap E(x_n,D_n))=\mu(E(y_{-n}^i;\{x^i\}_{-n};x_n,D_n))=m(y_{-n}^i;\{x^i\}_{-n};x_n,D_n)$ and $\mu(C(x,A)^{-n})=\mu(C(x_{-n},A_{-n};x_n,\{x_n\})=p(x_{-n},A_{-n};x_n,\{x_n\})=p_{-n}(x_{-n},A_{-n})$.

($\impliedby$): Since $|X_t|=3$ for all $1\leq t<n$, 
\begin{align*}
    m(y_{-n},\{x\}_{-n})=\sum_{B_{-n}\geq \{x\}_{-n}^C} (-1)^{\sum_{t=1}^{n-1} |B_t|-2(n-1)} p_{-n}(y_{-n},B_{-n})\geq 0 \\ 
    \iff \sum_{B_{-n}\geq \{x\}_{-n}^C: \sum_{t=1}^{n-1} |B_t| \text{even}} p_{-n}(y_{-n},B_{-n}) \geq \sum_{B_{-n}\geq \{x\}_{-n}^C: \sum_{t=1}^{n-1} |B_t| \text{odd}} p_{-n}(y_{-n},B_{-n})
\end{align*}
and hence Partial Marginal BM Nonnegativity is equivalent to Joint Supermodularity of $p_{-n}$. By definition, $(-n)$-Marginal Consistency is equivalent to Marginal Consistency of $p_{-n}$. Hence, by \textbf{Theorem \ref{p:all3}}, $\rho_{-n}$ has a unique SU representation $\mu_{-n} \in \Delta(P_{-n})$.

Next, fix any $y_{-n}\neq\neq x_{-n} \in X_{-n}$ with $m(y_{-n},\{x\}_{-n})>0$, and identify $(y_{-n},\{x\}_{-n})$ with the (unique) $\succ_{-n}$ such that $\{\succ_{-n}\}=E_{-n}(y_{-n},\{x\}_{-n})$. For any $x_n \in A_n\in \mathcal{M}_n$, \textbf{Lemma \ref{l:1perBMsum}} and Partial Conditional BM Nonnegativity imply
$$\rho_n(x_n,A_n|(y,\{x\})^{-n})=\sum_{B_n\subseteq A_n^C} m(x_n,B_n|(y,\{x\})^{-n})\geq 0$$ 
Furthermore, by \textbf{Lemma \ref{l:condSCFBM}},
$$\sum_{x_n \in A_n} \rho_n(x_n,A_n|(y,\{x\})^{-n})=1$$
Hence, $\rho_n(\cdot|(y,\{x\})^{-n})$ is an SCF. By \cite{block1959random}'s characterization of static RU, Partial Conditional BM Nonnegativity implies that $\rho_n(\cdot|(y,\{x\})^{-n})$ has an RU representation $\mu^{\succ_{-n}} \in \Delta(P_n)$. For $\succ_{-n}$ with $\mu_{-n}(\succ_{-n})=0$, define $\mu^{\succ_{-n}} \in \Delta(P_n)$ to be any fixed probability measure over $P_n$.

Next, define $\mu: 2^P\rightarrow \R$ as
$$\mu(\succ):=\mu_{-n}(\succ_{-n})\mu^{\succ_{-n}}(\succ_n)$$
and $\mu(C):=\sum_{\succ \in C} \mu(\succ)$ for all (non-singleton) $C \in 2^P$. By definition of $\mu$, $\mu\geq 0$ and
$$\sum_{\succ \in P} \mu(\succ)=\sum_{\succ_{-n} \in P_{-n}} \mu_{-n}(\succ_{-n}) \sum_{\succ_n \in P_n}\mu^{\succ_{-n}}(\succ_n)=\sum_{\succ_{-n} \in P_{-n}} \mu_{-n}(\succ_{-n})=1$$
Hence, $\mu$ is a probability measure over $P$. More generally, for any $C_{-n}\subseteq P_{-n}$ and $C_n\subseteq P_n$,
$$\mu(C_{-n} \times C_n)=\sum_{\succ_{-n} \in C_{-n}} \mu_{-n}(\succ_{-n})\sum_{\succ_n \in C_n} \mu^{\succ_{-n}}(\succ_n)=\sum_{\succ_{-n} \in C_{-n}} \mu_{-n}(\succ_{-n}) \mu^{\succ_{-n}}(C_n)$$
Hence, for any $x_1 \in A_1 \in \mathcal{M}_1$,
\begin{align*}
    \mu(C(x_1,A_1))=\mu(C_{-n}(x_1,A_1) \times P_n)=\sum_{\succ_{-n} \in C_{-n}(x_1,A_1)} \mu_{-n}(\succ_{-n}) \mu^{\succ_{-n}}(P_n) \\
    =\mu_{-n}(C_{-n}(x_1,A_1))=\rho_1(x_1,A_1)
\end{align*}
since $\mu_{-n}$ is an SU representation of $\rho_{-n}$. Similarly, fix any $1<t<n$ and let $T=\{1,\ldots,t\}$, $T-1=T\backslash\{t\}$. Fix any $(A,x)^{T-1} \in \mathcal{H}_{t-1}$ and $x_t \in A_t \in \mathcal{M}_t$: then
\begin{align*}
    \mu(C(x_t,A_t)|C(A,x)^{T-1})=\frac{\mu(C(x,A)^T)}{\mu(C(x,A)^{T-1})}=\frac{\mu(C_{-n}(x,A)^T \times P_n)}{\mu(C_{-n}(x,A)^{T-1} \times P_n)} \\
    =\frac{\mu_{-n}(C_{-n}(x,A)^T)}{\mu_{-n}(C_{-n}(x,A)^{T-1})}=\frac{p_{-n}(x,A)^T}{p_{-n}(x,A)^{T-1}}=\frac{p_t(x,A)^T}{p_{t-1}(x,A)^{T-1}}=\rho_t(x_t,A_t|(A,x)^{T-1})
\end{align*}
where the fourth equality holds because
$\mu_{-n}$ being an SU representation of $\rho_{-n}$ and $C_{-n}(x,A)^T=C_{-n}(x_T,A_T;x_{-n\backslash T},\{x\}_{-n\backslash T})$ imply $\mu_{-n}(C_{-n}(x,A)^T)=p_{-n}(x_T,A_T)$, and the fifth equality holds because $p_t$ and $p_{t-1}$ are the marginals of $p_n$ on $A_T$ and $A_{T-1}$, respectively.

Finally, for any $(A,x)^{-n} \in \mathcal{H}_{n-1}$ and $x_n \in A_n \in \mathcal{M}_n$, let $\{\succ_{-n}^i\}_{i \in I}$ be the unique singleton partition of $(A,x)^{-n}$ and let $I':=\{i \in I: \mu_{-n}(\succ_{-n}^i)>0\}$. Identify $\succ_{-n}^i$ with $(y_{-n}^i,\{x^i\}_{-n})$ as before, and define $(\succ_{-n}^i)$ as before. Then
\begin{align*}
    \mu(C(x_n,A_n)|C(A,x)^{-n})=\sum_{i \in I'} \mu(C(x_n,A_n)|\succ_{-n}^i)\mu(\succ_{-n}^i|C(x,A)^{-n}) \\
    =\sum_{i \in I'} \frac{\mu(C(x_n,A_n)\cap (\succ_{-n}^i))}{\mu(\succ_{-n}^i)} \frac{\mu(\succ_{-n}^i)}{\mu(C(x,A)^{-n})} \\
    =\sum_{i \in I'} \frac{\mu(\{\succ_{-n}^i\}\times C_n(x_n,A_n))}{\mu(\succ_{-n}^i)}\frac{\mu(\succ_{-n}^i)}{\mu(C_{-n}(x,A)^{-n}\times P_n)} \\
    =\sum_{i \in I'} \frac{\mu_{-n}(\succ_{-n}^i)\mu^{\succ_{-n}^i}(C_n(x_n,A_n))}{\mu_{-n}(\succ_{-n}^i)}\frac{\mu_{-n}(\succ_{-n}^i)}{\mu_{-n}(C_{-n}(x,A)^{-n})} \\
    =\sum_{i \in I'} \frac{\sum_{D_n\subseteq A_n^C}\mu_{-n}(\succ_{-n}^i)\mu^{\succ_{-n}^i}(E_n(x_n,D_n))}{\mu_{-n}(\succ_{-n}^i)}\frac{\mu_{-n}(\succ_{-n}^i)}{\mu_{-n}(C_{-n}(x,A)^{-n})} \\
    =\sum_{i \in I'} \frac{\sum_{D_n\subseteq A_n^C} m(y_{-n}^i,\{x^i\}_{-n})m(x_n,D_n|y_{-n}^i,\{x^i\}_{-n})}{m(y_{-n}^i,\{x^i\}_{-n})}\frac{m(y_{-n}^i,\{x^i\}_{-n})}{p_{-n}(x_{-n},A_{-n})} \\
    =\sum_{i \in I'} \rho_n(x_n,A_n|(y^i,\{x^i\})^{-n})\frac{m(y_{-n}^i,\{x^i\}_{-n})}{p_{-n}(x_{-n},A_{-n})}=\rho_n(x_n,A_n|(A,x)^{-n})
\end{align*}
where the fifth equality holds because $C_n(x_n,A_n)=\bigcup_{D_n\subseteq A_n^C} E_n(x_n,D_n)$ is disjoint, the sixth equality holds by applying \textbf{Proposition \ref{p:SU_iff_assigns}} to $(\rho,p,\mu)_{-n}$ and the static analog of \textbf{Proposition \ref{p:SU_iff_assigns}} to $\rho_n(\cdot|y_{-n}^i,\{x^i\}_{-n})$ and $\mu^{y_{-n}^i,\{x^i\}_{-n}}$ for each $i \in I'$,\footnote{For the static analog of \textbf{Proposition \ref{p:SU_iff_assigns}}, see Proposition 7.3 of \cite{chambers2016revealed}.} the seventh equality holds by \textbf{Lemma \ref{l:condSCFBM}} and by definition of $\rho_n(\cdot|y_{-n}^i,\{x^i\}_{-n})$, and the eighth equality holds by $(-n)$-Conditional Consistency.
\end{proof}

\subsection{Proof of Theorem \ref{t:jcoh}}
\begin{proof}
($\implies$): Let $\mu$ be an SU representation, and fix any $(x^i,A^i)_{i=1}^k$ with $x^i \in A^i \in \mathcal{M}$ for each $1\leq i\leq k$ and any $(\lambda^i)_{i=1}^k\subseteq \R$ such that $\sum_{i=1}^k \lambda^i \mathbbm{1}_{C(x^i,A^i)}\geq 0$. Hence,
\begin{align*}
    \sum_{i=1}^k \lambda^i p(x^i,A^i)=\sum_{i=1}^k \lambda^i \mu(C(x^i,A^i))=\mathbb{E}_\mu\bigg[\sum_{i=1}^k \lambda^i \mathbbm{1}_{C(x^i,A^i)}\bigg]\geq 0
\end{align*}
where the first equality follows from \textbf{Proposition \ref{p:SU_iff_assigns}}, the second equality follows from linearity of expectation and because the expectation of an indicator random variable of an event is the probability of that event, and the inequality follows because a convex combination of nonnegative numbers is nonnegative.

($\impliedby$): Using the notation from \cite{clark1996random}, let $\mathcal{A}:=\{C(x,A)\}_{x \in A \in \mathcal{M}}$ and note that $P=C(x,\{x\}_N) \in \mathcal{A}$. Let $\hat{\mathcal{A}}$ be the algebra generated by $\mathcal{A}$. First, I show $\hat{\mathcal{A}}=2^P$. To show this, it suffices to show that $\hat{\mathcal{A}}$ contains all singletons, since $\hat{\mathcal{A}}$ is closed under finite unions and every event in $2^P$ is a finite union of singletons. Fix any $\succ \in P$ and identify it with its ranking of all elements in each period: $(x_t^1,\ldots,x_t^{|X_t|})_{t=1}^n$. Define indices $t_1,\ldots,t_n$ such that $|X_{t_1}|\leq \cdots \leq |X_{t_n}|$. Hence,
\begin{align*}
    \{\succ\}=\bigcap_{k=1}^{|X_{t_n}|} C(x^k,\{x_t^k,\ldots,x_t^{|X_t|}\}_N) \in \hat{\mathcal{A}}
\end{align*}
where $x_t^k:=x_t^{|X_t|}$ and $\{x_t^k,\ldots,x_t^{|X_t|}\}:=\{x_t^{|X_t|}\}$ for $k>|X_t|$, and since $\hat{\mathcal{A}}$ is closed under finite intersections.\footnote{For example, let $n=2$ and $\succ=(x_1^1x_1^2x_1^3,x_2^1x_2^2x_2^3x_2^4)$. Then $\{\succ\}=C(x^1,X)\cap C(x_1^2,x_1^3;x_2^2,\{x_2^2,x_2^3,x_2^4\}) \cap C(x_1^3,\{x_1^3\};x_2^3,x_2^4\} \cap C(x_1^3,\{x_1^3\};x_2^4,\{x_2^4\})$.} Consider $p: \mathcal{A}\rightarrow \R$ as $p(C(x,A)):=p(x,A)$ and note that $p(x,\{x\}_N)=1$. By Theorem 1 of \cite{clark1996random}, there exists a finitely additive probability measure $\mu: 2^P \rightarrow [0,1]$ such that $\mu(C(x,A))=p(x,A)$ for all $x \in A \in \mathcal{M}$. By \textbf{Proposition \ref{p:SU_iff_assigns}}, $\mu$ is a SU representation of $\rho$.
\end{proof}

\bibliography{BibFile}

\end{document}